\documentclass[prd,aps,nofootinbib]{revtex4} %showpacs,

\usepackage{latexsym, graphicx}
\usepackage{color, hyperref}

\begin{document}
%\draft

\title{Notes on observational and radar coordinates for localized observers}
\author{Shih-Yuin Lin}
\email{sylin@cc.ncue.edu.tw}
\affiliation{Department of Physics, National Changhua University of Education, Changhua 50007, Taiwan}

\date{\today}

\begin{abstract}
The worldline of a uniformly accelerated localized observer in Minkowski space is restricted in the Rindler wedge, where the observer can in principle arrange experiments repeatedly, and the Cauchy problem for quantum fields in that Rindler wedge can be well defined. However, 
the observer can still receive the signals sourced by the events behind the past horizon, and coordinatize those events in terms of
some kind of observational coordinates. We construct such observational coordinates in some simple cases with the localized observers in Minkowski, de Sitter, and Schwarzschild-like spacetimes, and compare them with radar coordinates for the same observers. 
\end{abstract}

%\pacs{PACS numbers: }
\maketitle
 
%%%%%%%%%%%%%%%%%%%%%%%%%%%%%%%%%%%%%%%%%%%%%%%%%%%%%%%%%%%%%%%%%%%%%%%%%%%%%%%%%%%%%%%%%%%%%%%%%%%%
\section{Introduction}

In special relativity, an observer is considered to be localized in space with a clock \cite{Ei05}\footnote{
The ``local" observer in general relativity refers to the one confined to a finite region where the variation of gravitational field is unobservably small \cite{DIn92, Ma90}. %p.129
In addition to this locality, the ``localized" observer considered in this paper has infinitesimal volume to ensure causality and reduce the ambiguity of the coordinates determined by her.}. 
Such a localized observer moving at a constant velocity in Minkowski space can operationally define a reference frame, 
called ``radar coordinates", for the events in spacetime by sending a radar pulse at her proper time $\tau_i$ to some event and then recording the receiving time $\tau_f$ of the echo from the event. Accordingly, each event can be coordinatized in terms of radar time $t=(\tau_f+\tau_i)/2$, radar distance $r=(\tau_f-\tau_i)/2$, and the direction of sending/receiving the radar signal. The principle of special relativity (encoded in Bondi's $k$-calculus) \cite{Bo65, MTW73, DIn92} implies that the radar coordinates constructed in this way coincide with the Minkowski coordinates Lorentz transformed from those for a rest observer.

\begin{figure}
\includegraphics[width=5.5cm]{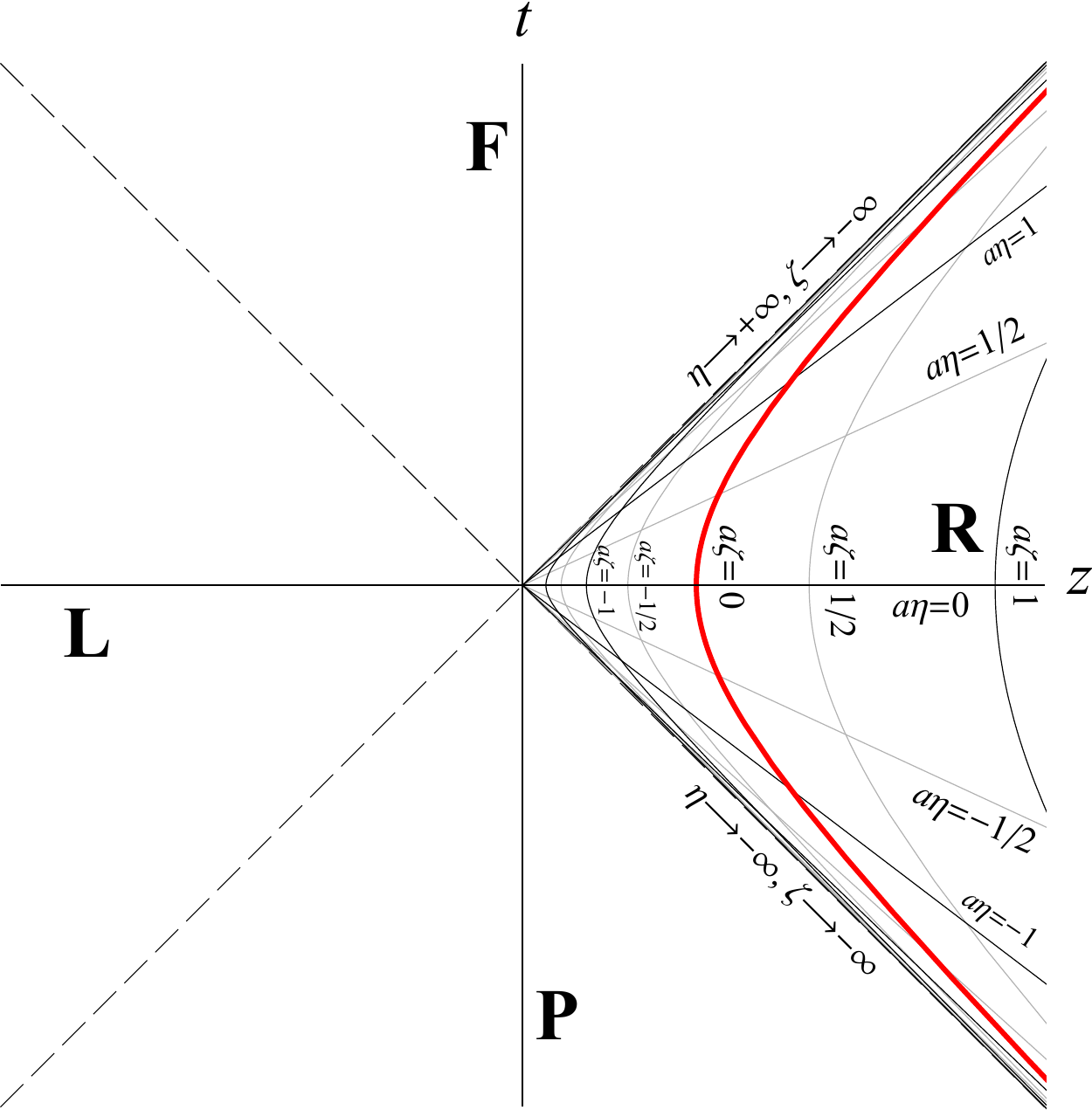} \hspace{.5cm}
\includegraphics[width=5.5cm]{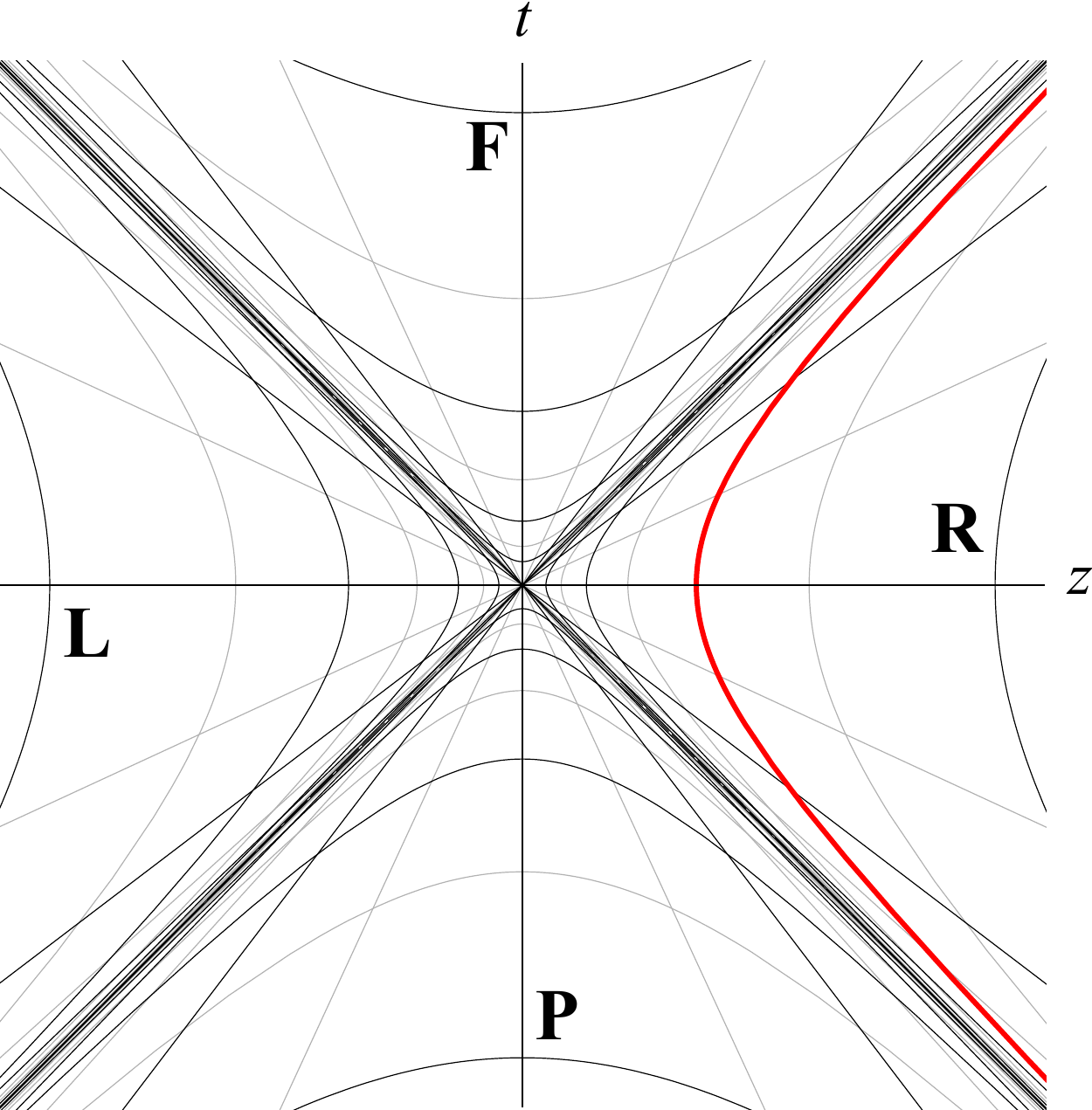} %{RindlerMax_tz.pdf}
\caption{Rindler coordinates in Minkowski space (left) and its maximal analytic extension (right).}
\label{Rind}
\end{figure}

A similar coordinatization scheme also works for an accelerated localized observer with an ideal clock unaffected by its acceleration 
\cite{DIn92, DP87, PV00, DG01} if the acceleration is not too large to invalidate the hypothesis of locality \cite{Ma90}. In particular, for a uniformly accelerated observer at proper acceleration $a$ in (1+1)D Minkowski space, radar coordinates are exactly the conventional Rindler coordinates \cite{La63, Pa21, Ri66} 
\begin{equation}
  ds^2 = e^{2a\zeta}\left(-d\eta^2 + d\zeta^2\right),  \label{Rind2D}
\end{equation}
defined only in the wedge containing the observer's worldline (wedge R in Figure \ref{Rind}), with radar time $\eta \in (-\infty, \infty)$, radar distance $|\zeta|\in [0,\infty)$, and the directions of the radar signals indicated by the sign of $\zeta$. 
There are some advantages in applying Rindler radar coordinates to field theory, e.g.
a Lorentz boost about the origin in the Minkowski coordinates is simply a time translation in Rindler radar coordinates, and the Cauchy problem for quantum fields can be well defined in Rindler radar coordinates.
Nevertheless, the uniformly accelerated observer in wedge R should be able to receive the classical signal emitted from an event behind the past horizon of radar coordinates such as the outcome of a local measurements in wedge P on a field. How could the uniformly accelerated observer coordinatize that event, which is never reachable by her radar signals?

When we look into the sky, we are receiving the information along the past light cones extended from our eyes. 
Astronomers can see the classical signals emitted by an object billions of light years away from the Earth, 
while radar was invented only decades ago. 
%The visible universe is much larger than the region that can be practically covered by any radar coordinates. 
To coordinatize an observed event beyond the reach of any radar signal, %coordinates,
one may follow astronomers and adopt ``optical coordinates" \cite{Te38}, also called ``observational coordinates" \cite{El85} or ``geodesic light-cone coordinates" \cite{Nu11, Nu16}, in terms of the signal-receiving time in the observer's clock together with the distance and the direction of the event seen by the observer.
%The observational coordinates could uniquely identify the events in the visible universe without too strong gravitational lensing of compact objects. 

In (3+1)D, the astronomical distances which may be useful for observational coordinates include the binocular (parallax) distance, luminosity distance, angular diameter distance, and so on \cite{We72}. In the ideal cases the affine distance and advanced/retarded distance by mathematical constructions would also be convenient for theorists \cite{HP10}. To determine the astronomical distance of an object at some moment, either the observer or the observed object has to be extended in the directions orthogonal to the null geodesic connecting the object and the observer at that moment, while the sizes of the observer and the observed object/event are considered to be infinitesimal in this paper. Indeed, the angular diameter distance of an object cannot be determined if its angular diameter is zero, and one needs a baseline between two eyes or telescopes to determine the binocular distance, or an antenna of finite area for measuring the apparent luminosity to determine the luminosity distance from an event.
In (1+1)D, however, while radar distance can still be defined, those astronomical distances cannot be determined physically 
and we have to rely on the distances by mathematical constructions to depart from radar coordinates.

The idea of the observational coordinates is not new. Similar ideas have been applied to curved spacetimes to construct 
the advanced coordinates, which are the time reversal of the retarded coordinates \cite{NU63, Po03}.
And yet, some details of observational coordinates and even radar coordinates are not fully explored.
What would the events behind the past horizon of Rindler radar coordinates look like in the viewpoint of a uniformly accelerated localized observer? Does an accelerated localized observer always see a past horizon of radar coordinates and the events behind it? 
There is no nontrivial coordinate singularity at finite values of Rindler radar coordinates (\ref{Rind2D}). How about observational coordinates? Is the acceleration of the localized observer necessary for the presence of nontrivial coordinate singularities?

To answer these questions, below we study a few simple cases of the localized observers in various motions and background geometries.
In section \ref{ALOMink} we look at the localized observer in uniform acceleration, non-uniform acceleration, and spinning without center-of-mass motion in (3+1)D Minkowski space. The cases with a comoving localized observer in de Sitter space and in a non-eternal inflation background are discussed in section \ref{StatOdeS}. Then the case with the localized observer fixed at a constant radius from the center of a spherical shell in (1+1)D is considered in section \ref{Schwarz2D}. Finally, we summarize our findings with discussion in section \ref{sumdisc}.

\section{Localized observers in Minkowski space}
\label{ALOMink}

\begin{figure}
\includegraphics[width=5.5cm]{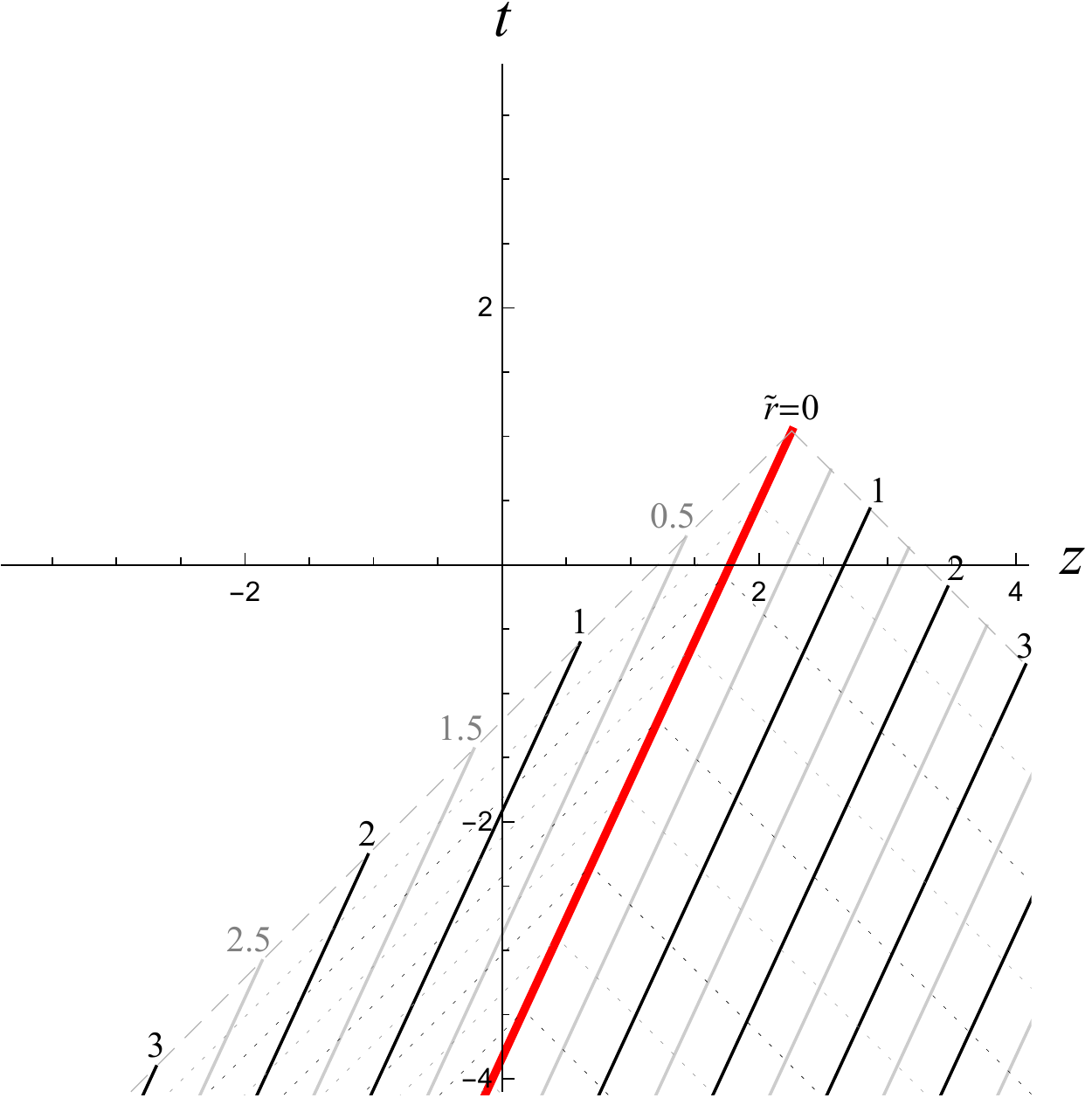} %{POVU_tz.pdf}
\includegraphics[width=5cm]{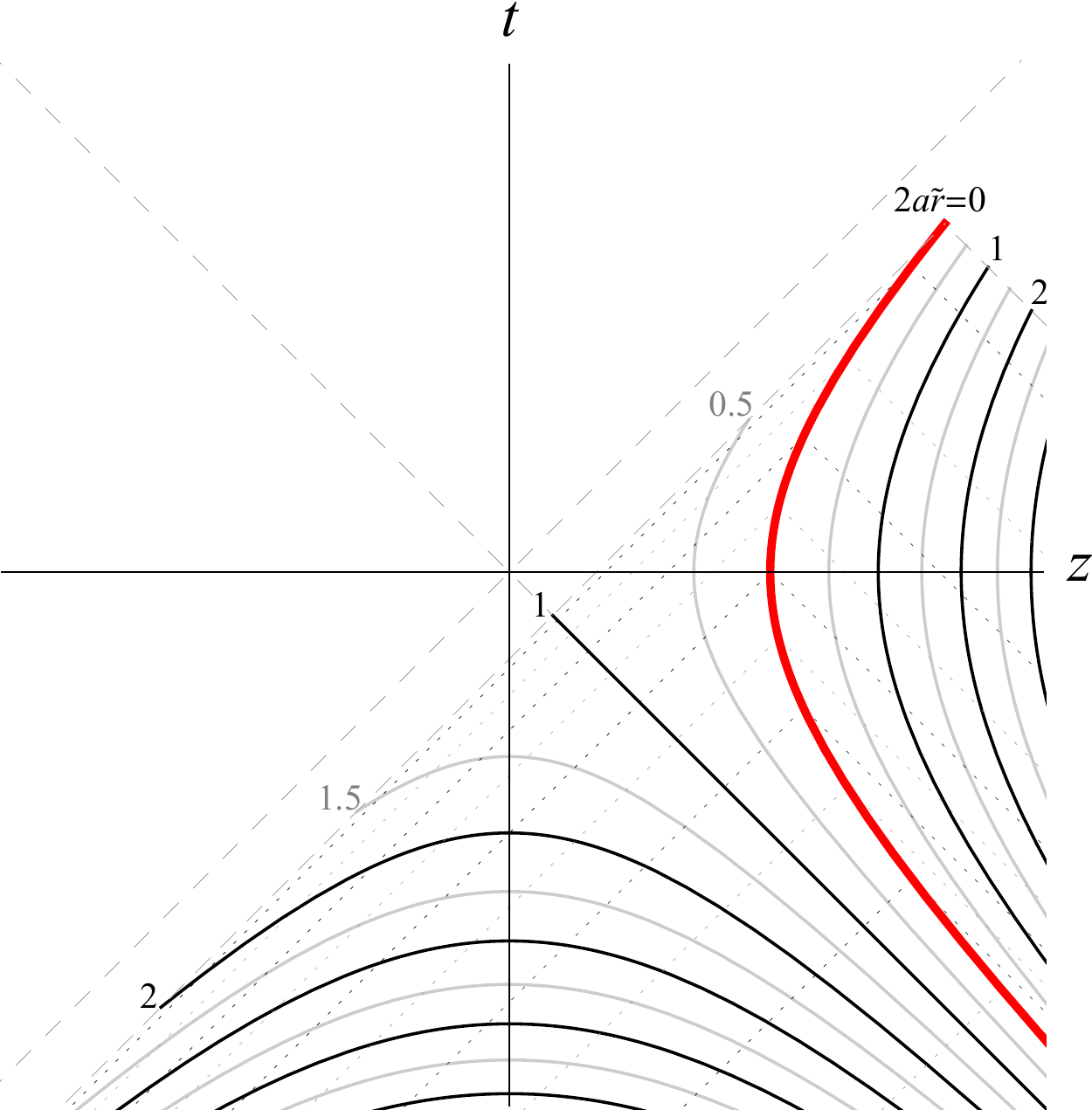} %{POVUA_tz.pdf}
\includegraphics[width=5.5cm]{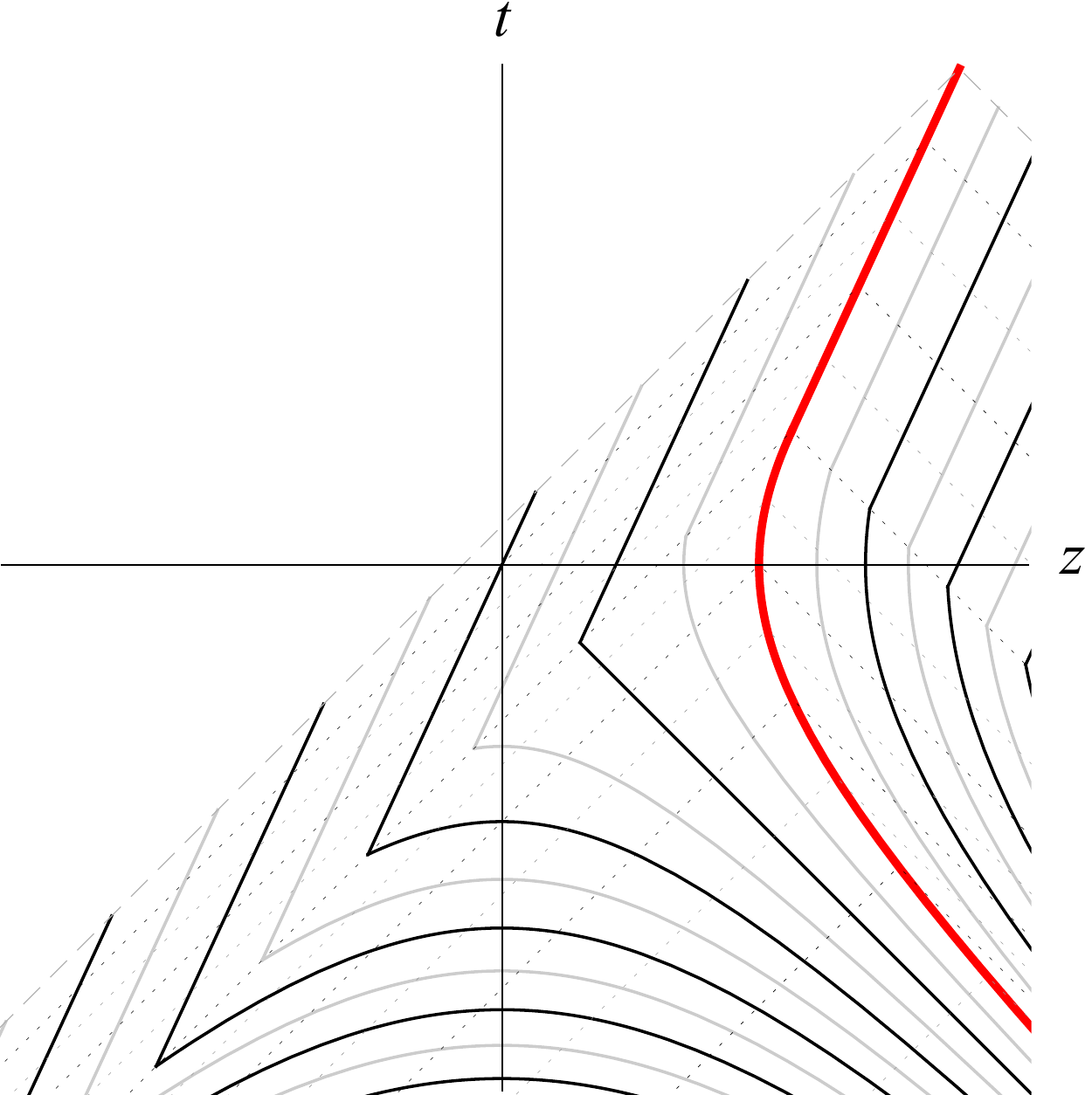} %{POVUAU_tz.pdf}
\caption{The $x^0 x^3$-plane in Minkowski coordinates now coordinatized by an observer in uniform motion (left), uniform acceleration (middle), and constant velocity after a period of constant linear acceleration (right) in the observational coordinates. Here $x^0$ and $x^3$ are denoted by $t$ and $z$, respectively. The red curves represent the worldlines of the observer, the black and gray solid lines represent the constant $\tilde{r}$ (advanced distance to the localized observer) hypersurfaces, and the black and gray dotted lines
represent the constant $\tilde{\tau}$ (proper time of the localized observer) hypersurfaces. The black dashed lines in the middle plot represent the hypersurfaces $x^0-x^3=0$ and $x^0+x^3=0$, which are the event horizon and the past horizon for the uniformly accelerated observer, respectively.}
\label{POVMin}
\end{figure}

The observational coordinates with the observer's clock $\tilde{\tau}$, the distance $\tilde{r}$, and the direction $(\tilde{\theta}, \tilde{\varphi})$ of the event seen by a non-spinning localized observer in inertial motion in Minkowski space will coincide with the radar coordinates for the same observer and the conventional Minkowski coordinates up to a Lorentz transformation if we define the time coordinate as $t=\tilde{\tau}-\tilde{r}$ (Figure \ref{POVMin} (left)). 
Similar observational coordinate systems can also be determined for localized observers in general motions and background spacetimes, but they usually deviate from radar or other conventional coordinates for the same observers.

%%%%%%%%%%%%%%%%%%%%%%%%%%%%%%%%%%%%%%%%%%%%
\subsection{Uniformly accelerated observer}
\label{UAO}

Consider the simplest case of non-inertial motions, where the localized observer with proper time $\tau$ is uniformly accelerated with proper acceleration $a$ along the worldline $z^\mu(\tau)=(a^{-1}\sinh a\tau, 0,0, a^{-1}\cosh a\tau)$ in Minkowski coordinates $x^\mu$ 
%= (t,x,y,z)$ 
in (3+1)D Minkowski space. For this uniformly accelerated localized observer, a conventional, natural choice of the coordinate system would be Rindler coordinates %$(\eta,x,y,\zeta)$ given by
\begin{equation}
  ds^2 = e^{2a\zeta}\left(-d\eta^2 + d\zeta^2\right) + (dx^1)^2 + (dx^2)^2, \label{Rind4D}
\end{equation}
with $-\infty < \eta < \infty$, $-\infty < \zeta < \infty$, and $x^1$ and $x^2$ being identical to those of Minkowski coordinates.
The metric (\ref{Rind4D}) is transformed from Minkowski coordinates $ds^2 = \eta_{\mu\nu}dx^\mu dx^\nu$, $\eta_{\mu\nu}={\rm diag}(-1,1,1,1)$ by $x^0 = a^{-1} e^{a\zeta} \sinh a\eta$ and $x^3=a^{-1} e^{a\zeta}\cosh a\eta$. The uniformly accelerated observer appears to be at rest along the worldline $(\tau,0,0,0)$ in Rindler coordinates $(\eta, x^1, x^2, \zeta)$. 
Unlike the case in (1+1)D, however, Rindler coordinates (\ref{Rind4D}) in (3+1)D is not a radar coordinate system as $\zeta$ is not the radar distance of any event off the plane of the observer's motion $x^1=x^2=0$. 

\subsubsection{Radar coordinates}

The radar coordinates for the uniformly accelerated observer going along $z^\mu(\tau)$ can be obtained 
%The line element (\ref{UAradar}) is obtained 
using the same operations as those in the inertial cases. %, as follows. 
Suppose the observer emitted a radar pulse at $\tau=\tau_i$ to an event $E$ at $x^\mu =x^\mu_E\equiv (t,x,y,z)$ in Minkowski coordinates and received the echo at $\tau=\tau_f$. Then, the radar time and radar distance for the event will be $\eta = (\tau_f+\tau_i)/2$ and $r = (\tau_f-\tau_i)/2$, respectively, and the event $E$ will be somewhere in the $\eta$-slice of $(\xi \sinh a\eta,x^1,x^2, \xi \cosh a\eta)$  
%, and $x^1, x^2 \in {\bf R}$ 
in terms of Minkowski coordinates ($\xi\in{\bf R}$).

\begin{figure}
\includegraphics[width=4.5cm]{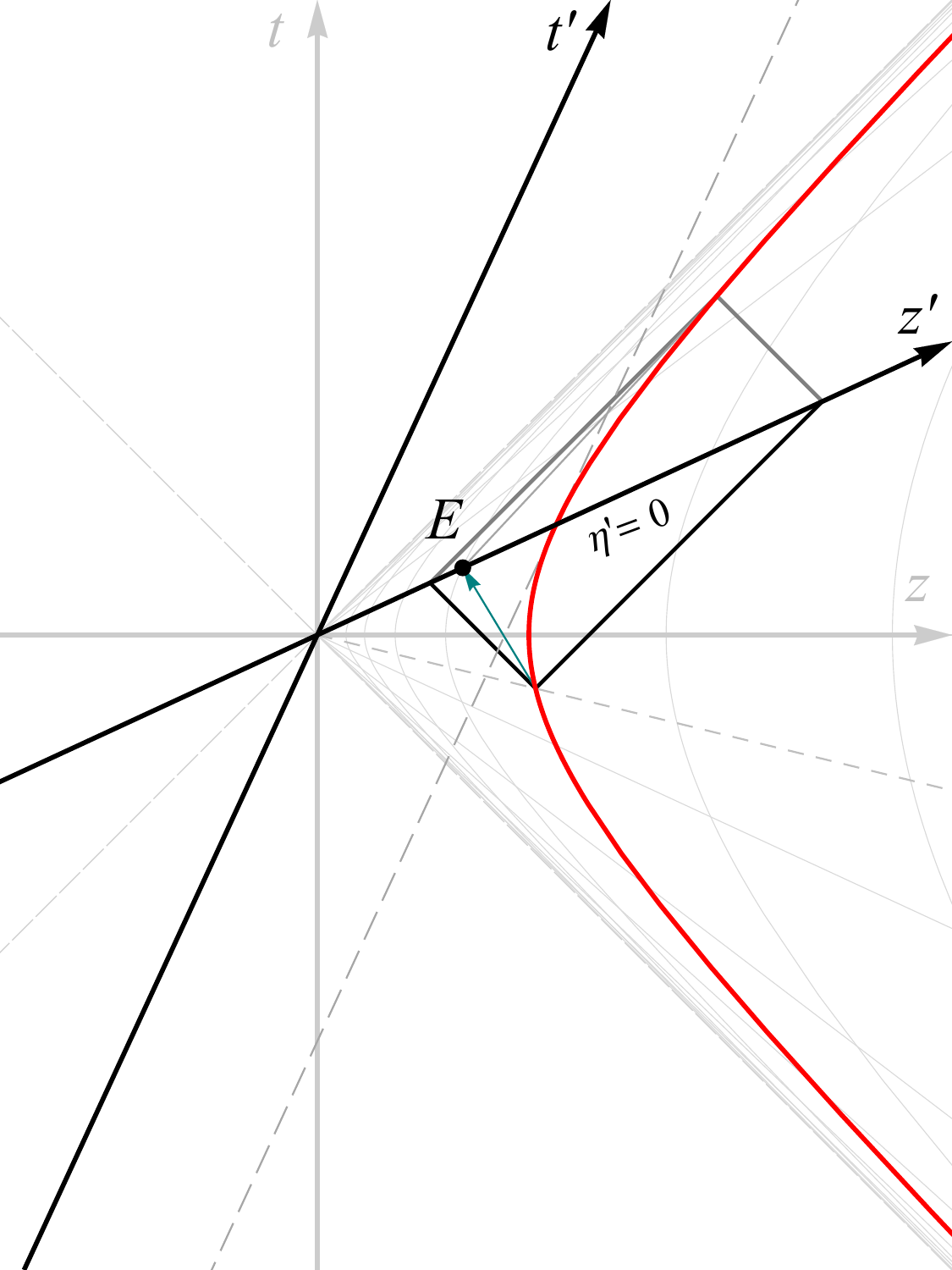} \hspace{.5cm}
\includegraphics[width=6cm]{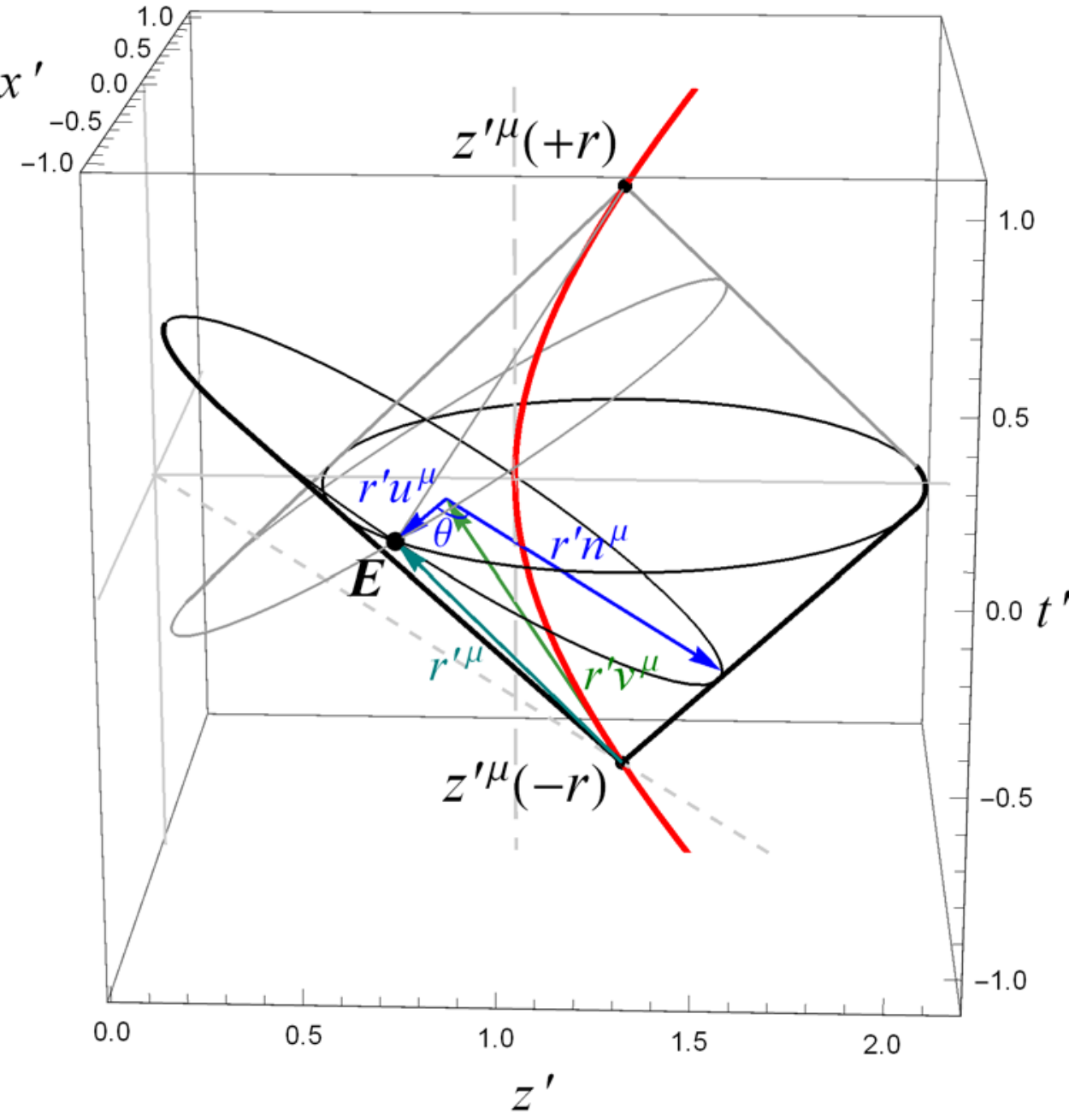} \hspace{.5cm}
\includegraphics[width=5.5cm]{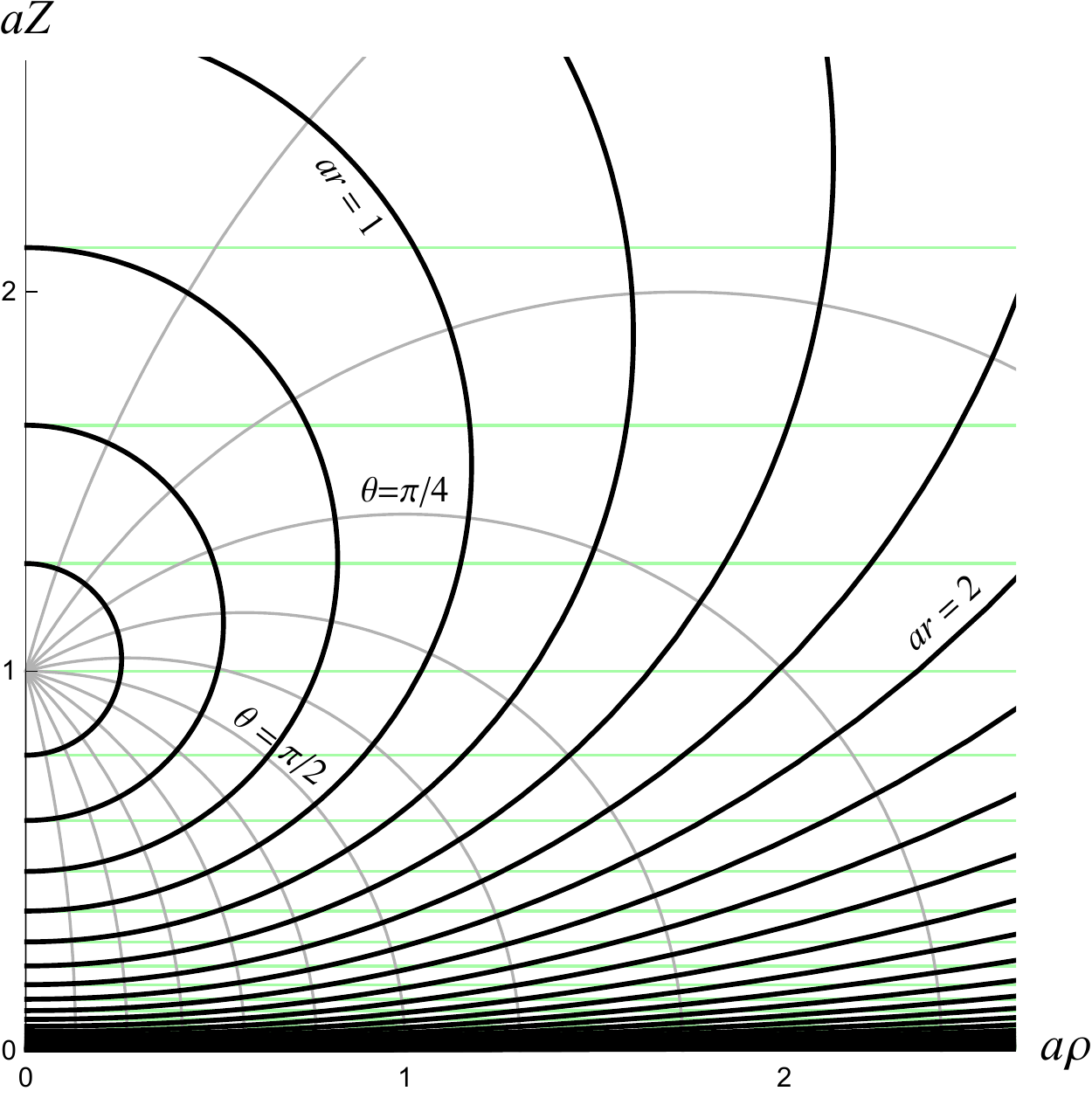} %{RadarUA.pdf}
\caption{(Left) Transform to a new coordinates $x'^\mu$ by a Lorentz boost so that the event $E$ is in the slice of $t'=\eta'=0$.
(Middle) In the new coordinates, let $r'^\mu = x'^\mu_E - z'^\mu(-r)$, where $x'^\mu_E$ represents the event $E$ and $z'^\mu(-r)$ is the event that the observer emits the radar pulse. Define the spacelike 4-vector $u^\mu$ by $r'^\mu \equiv r' (v^\mu + u^\mu)$, where $v^\mu$ 
is the 4-velocity of the observer when emitting the pulse. Let $n^\mu$ be the spacelike vector points to the direction of acceleration in view of the observer. Then the observer determines the angle of departure $\theta$ of the radar pulse by $u_\mu n^\mu = \cos \theta$. The same $\theta$ will be observed as the angle of arrival of the radar echo at $\tau=+r$.
(Right) The black and gray curves represent the contours of $ar$ and $\theta$, respectively, of radar coordinates (\ref{UAradar}) on the $\eta'=0$ slice in our working Minkowski coordinates $x'^\mu$ for the uniformly accelerated observer (with $\varphi$ suppressed). The green lines are the contours of $a\zeta$ in the Rindler coordinates (\ref{Rind4D}) for the same observer. }
\label{radareta0}
\end{figure}

To determine $(\varphi,\theta)$ at $\tau_i$ and $\tau_f$, namely, the angles of departure and arrival of the radar pulse in the observer's point of view,
%\footnote{When a localized observer receives a radar echo coming from event $E$ along a null geodesic, the angle of arrival of the radar signal is the (solid-)angular position of the radar image of event $E$ in the sky of the observer, which is determined by the tangent of the null geodesic at the observer's location. The angle of departure is a similar angular position in the sky of a point-like emitter, where the radar signal departing from the emitter is pointing to. Here the point-like emitter of the radar pulse is the localized observer herself at $\tau'=-r$. Examples in a (3+1)D spherically symmetric space can be found in \cite{Lin19b}.}, 
one may perform a Lorentz boost in the $x^3$-direction about the origin to transform the $\eta$-slice to the $\eta'=0$ (and so $t'=0$) hypersurface in the new coordinates (Figure \ref{radareta0} (left)), in terms of which the event $E$ is represented as $x'^\mu \equiv (t',x',y',z')=(0,x,y,Z)\equiv x'^\mu_E$ and the events of emitting the radar pulse ($\tau=\eta-r$ or $\tau'=-r$) and receiving the echo ($\tau'=+r$) %for the observer 
are represented as $z'^\mu(\pm r)=(a^{-1}\sinh(\pm ar), 0, 0, a^{-1}\cosh ar)$. 
%Our working coordinates $x'^\mu$ is not the instantaneous Lorentz frame at the emitting event.
Let $r'^\mu \equiv x'^\mu_E - z'^\mu(-r)$ be a 4-vector pointing from the emitting event of the observer to the event $E$. Since $x'^\mu_E$ is on the future light cone of $z'^\mu(-r)$, %the observer at $\tau =\eta-r$ or $\tau'=-r$, 
one has $r'^\mu r'_\mu =0$, or
\begin{equation}
  (Z - a^{-1}\cosh a r)^2 + \rho^2 - (a^{-1}\sinh a r)^2 =0,  \label{nullcond}
\end{equation} 
with $\rho^2\equiv x^2+y^2$.
The spatial position $r'u^\mu$ of the event $E$ in the local Lorentz frame of the observer at $\tau'=-r$ but represented in our working coordinates $x'^\mu$ is defined by $r'^\mu = r'(v^\mu+u^\mu)$, where $v^\mu = \partial_{\tau'} z'^\mu(\tau')|_{\tau'=-r} = (\cosh ar, 0,0,-\sinh ar)$ is the 4-velocity or time direction of the localized observer at $\tau'=-r$, and one has $v^\mu v_\mu=-1$, $u^\mu u_\mu=+1$, and $u^\mu v_\mu=0$ (see Figure \ref{radareta0} (middle)). Immediately, one can see that $u^\mu = (r'^\mu/r')-v^\mu$ with $r' = -v^\mu r'_\mu = Z \sinh ar$. Suppose the localized observer chooses the direction of acceleration as the $z$-axis (where $(\theta,\varphi)\equiv (0,0)$) in her local frame and let $n^\mu = a^\mu/a|_{\tau'=-r} = (-\sinh ar, 0, 0, \cosh ar)$ be the spacelike unit vector in that direction ($a^\mu v_\mu =\partial_{\tau'}(v^\mu v_\mu)/2=0$ for all $\tau'$). Then, the azimuthal angle of departure $\varphi$ of the radar pulse in the observer's frame can be determined as usual, 
\begin{equation}
   \tan \varphi = y/x, \label{defphi}
\end{equation}
while the polar angle of departure $\theta$ is given by 
%$\tan \varphi |_{\tau_f=r} = \tan \varphi |_{\tau_i=-r} = y/x \equiv \tan \varphi$, and
$\cos \theta = u_\mu n^\mu = (Z \cosh a r - a^{-1})/(Z \sinh a r)$,
which implies 
\begin{equation} 
  Z = \frac{a^{-1}}{\cosh a r - \cos \theta \sinh a r} > 0 \label{Zofrtheta}
\end{equation}
and, together with the null condition (\ref{nullcond}), 
\begin{equation}
  \sin\theta =\frac{\rho}{Z \sinh a r}. \label{sintheta}
\end{equation}
The angle of arrival of the radar echo perceived by the observer can be obtained simply by a time-reversal transformation of the above argument (see the gray light cones in Figure \ref{radareta0} (middle)). It is clear that the azimuthal and polar angles of arrival $(\varphi,\theta)$ are exactly the same as the above angles of departure for the uniformly accelerated observer. Thus $(\varphi, \theta)$ can be adopted straightforwardly as the angular part of radar coordinates here. Transforming from our working coordinates $x'^\mu$ back to $x^\mu$ and from (\ref{sintheta}), one obtains the relations
%Then one has the null condition $\sigma(x'^\mu_E, z'^\mu(\pm r))=0$, or
%since $x'^\mu_E$ is on the future and past light cones of the observer at $\tau =\eta-r$ and $\eta+r$, respectively , and 
%%Now in the instantaneous Lorentz frame of the observer at $\tau=\pm r$ with the Lorentz factor $\gamma = \cosh a r$, the 3-speed $v=\tanh (\pm ar)$, and the origin set to be the observer's spacetime point of emitting or receiving the radar signal, the event $E$ is located at $(Z \sinh (\pm ar), x, y, Z \cosh a r - a^{-1})$, implying that 
%and
%\begin{equation}
%  \tan \theta |_{\tau_f=r} = \tan \theta |_{\tau_i=-r} = \frac{\rho}{Z \cosh a r - a^{-1}} \equiv \tan \theta \label{tantheta}
%\end{equation}
%and 
%so the observer would perceive that both the directions of emitting and receiving the radar signal are $(\theta, \varphi)$. 
%Combining Eqs. (\ref{nullcond}) and (\ref{tantheta}) yields $\sin\theta = \rho/(Z \sinh a r)$ and $\cos\theta = (Z \cosh a r - a^{-1})/(Z \sinh a r)$, which gives $Z = [a(\cosh a r - \cos \theta \sinh a r)]^{-1} > 0$ and the transformation
\begin{equation}
  t= Z \sinh a \eta, \hspace{1cm} 
  z = Z \cosh a \eta,	\hspace{1cm}
	\rho = Z \sinh a r \sin\theta,     \label{Min2radar}
\end{equation} 
between Minkowski coordinates $(t, z, \rho,\varphi)$ or $(t,x,y,z)$ and radar coordinates $(\eta, r, \theta, \varphi)$ of the event $E$, 
with $\varphi$ in (\ref{defphi}) invariant under the transformation. 
%and other coordinates transformed as (\ref{Min2radar}) 
Representing $Z(r, \theta)$ as (\ref{Zofrtheta}), the line element $ds^2= -dt^2 + dz^2 +d\rho^2 + \rho^2 d\varphi^2$ in Minkowski cylindrical coordinates can then be transformed to %190301-6
\footnote{Similar radar coordinates have been obtained in \cite{BLM04} with a different choice of the polar angle.}
\begin{equation}
  ds^2 = \frac{1}{(\cosh a r -\cos \theta \sinh a r)^2}\left[ -d\eta^2 + dr^2 + \left(\frac{\sinh ar}{a}\right)^2\left( d\theta^2 + 
	\sin^2 \theta  d \varphi^2 \right) \right] \label{UAradar}
\end{equation}
%after some algebra, 
with the values of radar coordinates $\eta \in (-\infty,\infty)$, $r\in [0, \infty)$, $\theta\in [0,\pi]$, and $\varphi\in[0,2\pi]$. In Figure \ref{radareta0} (right), we show the contours of constant $r$ (given by $\cosh ar = [(aZ)^2+(a\rho)^2+1]/(2aZ)$ from (\ref{nullcond})) and constant $\theta$ (given by $\tan\theta= a\rho/(aZ \cosh ar-1)=2a\rho/[(aZ)^2+(a\rho)^2-1]$ from (\ref{sintheta}) and (\ref{nullcond})) on the $\eta'=0$ slice in our working Minkowski coordinates $x'^\mu$ with $\varphi$ suppressed. It is clear that the above radar coordinates (\ref{UAradar}) for the uniformly accelerated observer coincide Rindler coordinates (\ref{Rind4D}) only in the plane of $\theta=0$ and $\pi$, i.e. $x^1=x^2=\rho=0$.

Both Rindler coordinates (\ref{Rind4D}) and radar coordinates (\ref{UAradar}) cover wedge R (Figure \ref{Rind} (left)) with $U\equiv x^0-x^3 (= -a^{-1}e^{a(\zeta-\eta)} = -e^{-a\eta}Z)<0$ and $V\equiv x^0+x^3(=a^{-1}e^{a(\zeta+\eta)}= e^{a\eta}Z)>0$ in Minkowski coordinates. Each spacetime point in wedge R is in principle accessible by the localized observer using those radar or light pulses, and so causally connected with the observer both in the past and future directions. The hypersurface $U=0$ is the event horizon and $V=0$ is the past horizon for the uniformly accelerated localized observer \footnote{Hereafter our ``past horizon for an observer" refers to the past horizon of the spacetime region covered by the radar coordinates for that observer. %, namely, the past boundary of the spacetime region that all the radar signal emitted by the observer during the whole history of the universe can possibly reach. 
No null geodesic starting at the observer can go beyond her past horizon, to which the observer would infer a divergent radar distance.}. 

The events behind the past horizon cannot be reached by any radar pulse from the uniformly accelerated observer and so
radar coordinates are not defined around those events. Nevertheless, the localized observer can passively receive the signal emitted by an event in wedge P and then determine the distance, which can be finite, from the emission event to the observer. Using the receiving time in the observer's clock, the distance, and the direction of the event perceived by the observer,
the localized observer can still coordinatize that event along the observer's past light cone.

\subsubsection{Observational coordinates}

Among the distances determined in different ways, the advanced distance may be the most convenient one in Minkowski space for theorists.
It can be read off from the field amplitude of a massless scalar field emitted by a point source as a standard candle \cite{Ro65, LH06, OLMH12}. Mathematically, the advanced distance of an event seen by the localized observer at some moment can be obtained by extrapolating the local Lorentz frame around the observer at that moment all the way to the event. While the advanced distance is not identical to the radar distance for non-inertial localized observers, for a general observer motion in (3+1)D Minkowski space it does coincide with the binocular distance, luminosity distance, angular diameter distance, proper-motion distance, as well as the affine distance of a null geodesic \cite{We72}. Below, we are using the advanced distance and the observer's proper time to construct an observational coordinate system explicitly for our uniformly accelerated observer. 

\begin{figure}
\includegraphics[width=6cm]{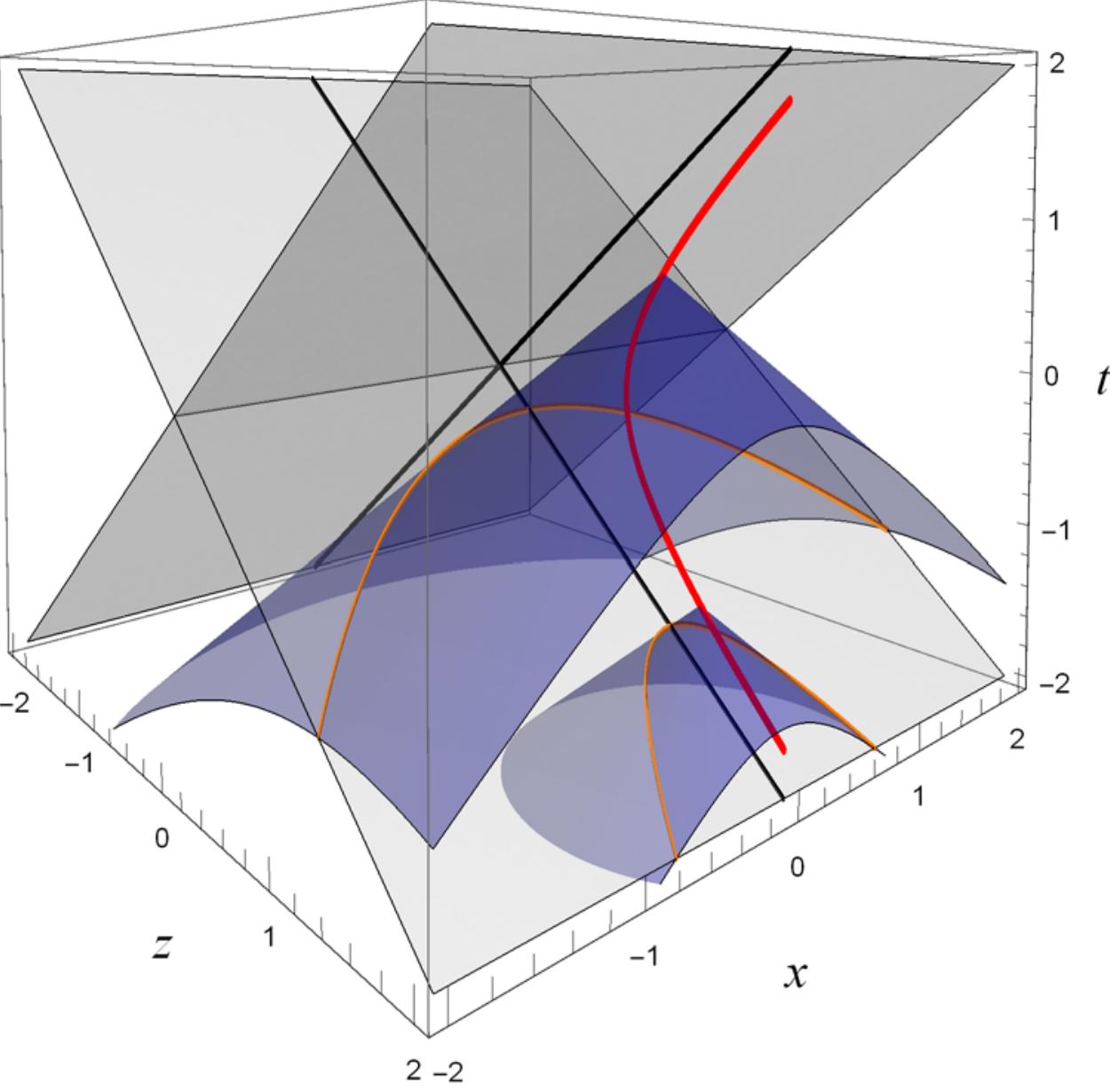} %{PastLC.pdf}
\includegraphics[width=8cm]{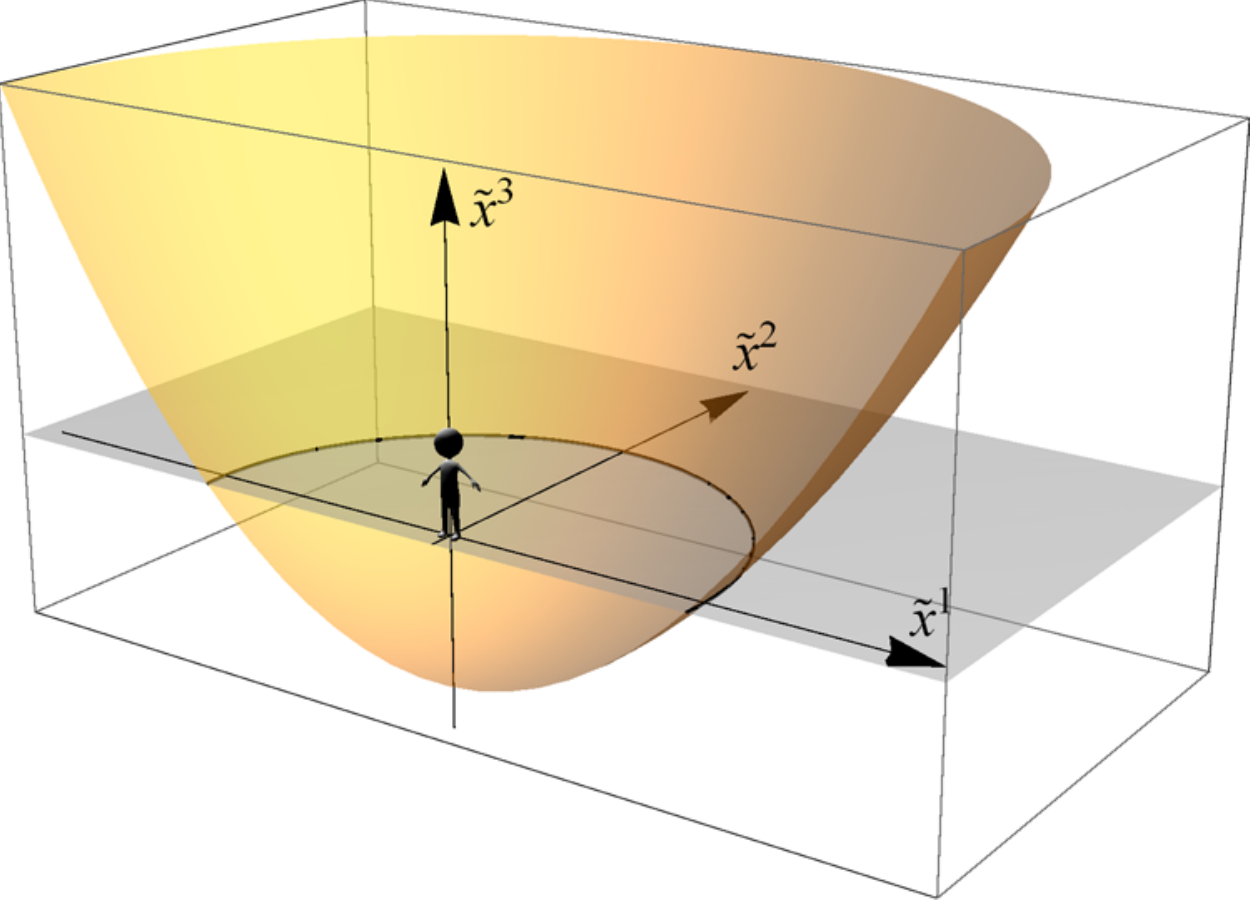} %{ObsHorizon.pdf}
\caption{(Left) The intersections of past light cones (blue) of the uniformly accelerated observer at different $\tilde{\tau}$ (red, $a=1$) and her past horizon $t=-z$ (gray) look like $\tilde{\tau}$-varying parabolas (orange) in Minkowski coordinates with the azimuthal angle $\tilde{\varphi}$ suppressed.  (Right) In fact, the past horizon is a {\it static} paraboloid of revolution (\ref{pasthor}) in the point of view of the uniformly accelerated observer at the origin of observational coordinates when $\tilde{\varphi}$ is shown.}
\label{intersec}
\end{figure}

Suppose a point source emits a light signal at $x^\mu=x^\mu_E\equiv(t,x,y,z)$ in the region of $U<0$ in Minkowski coordinates (wedges R and P in the maximally extended Rindler coordinates). The reading $\tilde{\tau}$ of the clock of the uniformly accelerated observer at the moment of receiving the light signal from the event $x^\mu_E$ is determined by the null condition $\sigma\left(z^\mu(\tilde{\tau}),x^\mu_E\right)=0$, where $\sigma(A^\mu, B^\mu) \equiv -\left(A_\mu -B_\mu\right) \left(A^\mu -B^\mu\right)/2$ is the Synge's world function. 
This condition gives 
$\tilde{\tau} = %\frac{1}{a} \ln \frac{a}{2|U|}\left( X - UV +\rho^2 +\frac{1}{a^2}\right),
	a^{-1}\ln [ a ( X - UV +\rho^2 +a^{-2})/(2|U|) ]$
with $U \equiv t-z$, $V\equiv t+z$, $\rho^2\equiv x^2+y^2$, 
and $X\equiv [ (t^2-z^2-\rho^2+a^{-2})^2+4 a^{-2}\rho^2 ]^{1/2}$ \cite{Ro65, LH06, OLMH12}. 
At the moment $\tilde{\tau}$, the 4-velocity of the observer is $\dot{z}^\mu(\tilde{\tau}) = (\cosh a\tilde{\tau}, 0,0,\sinh a\tilde{\tau}) = (\gamma^{}_{\tilde{\tau}},0,0,\gamma^{}_{\tilde{\tau}} v^{}_{\tilde{\tau}})$; thus, $\gamma^{}_{\tilde{\tau}}=\cosh a\tilde{\tau}$ and $v^{}_{\tilde{\tau}}= \tanh a\tilde{\tau}$. Choosing this spacetime point of the observer when receiving the signal, $z^\mu(\tilde{\tau})$, as the origin, %an extrapolated 
the position of the emission event $E$ in the local Lorentz frame %with the instantaneous 4-velocity $\dot{z}^\mu(\tilde{\tau})$ of 
for the observer at $\tilde{\tau}$ %in Minkowski coordinates 
can be obtained by a Poincar\'e transformation,
\begin{eqnarray}
    \tilde{x}^0 &=& \gamma^{}_{\tilde{\tau}}\left[ t - z^0(\tilde{\tau}) - v^{}_{\tilde{\tau}} (z-z^3(\tilde{\tau})) \right],\label{xO0}\\
		\tilde{x}^3 &=& \gamma^{}_{\tilde{\tau}}\left[ z - z^3(\tilde{\tau}) - v^{}_{\tilde{\tau}} (t-z^0(\tilde{\tau})) \right],\label{xO3}\\
		\tilde{x}^1 &=& x^1 - z^1(\tilde{\tau}) = x, \hspace{.5cm} \tilde{x}^2 = x^2 - z^2(\tilde{\tau})=y. \label{xO12}
\end{eqnarray}
Then, one has $-\tilde{x}^0 = \tilde{r} \equiv \sqrt{ (\tilde{x}^1)^2 + (\tilde{x}^2)^2 + (\tilde{x}^3)^2} = a X/2 = |\partial_\tau \sigma(z(\tau),x)|_{\tau=\tilde{\tau}}$, which is the advanced distance from the emitting event $(\tilde{x}^0,\tilde{x}^1,\tilde{x}^2,\tilde{x}^3)$ to the origin of this new coordinates, that is, the observer at $\tilde{\tau}$ \cite{OLMH12}. 
Using the four parameters $(\tilde{\tau}, \tilde{x}^1, \tilde{x}^2, \tilde{x}^3)$ as observational coordinates, the observer can uniquely identify the event at $x^\mu_E$. To find the metric for these observational coordinates, one can re-arrange the above relations in an inverse Poincar\'e transformation,
\begin{eqnarray}
    t - z^0(\tilde{\tau}) &=& \gamma^{}_{\tilde{\tau}}\left(\tilde{x}^0  + v^{}_{\tilde{\tau}} \tilde{x}^3 \right)
		 =  \gamma^{}_{\tilde{\tau}}\left(-\tilde{r}  + v^{}_{\tilde{\tau}} \tilde{x}^3 \right),\nonumber\\
		z - z^3(\tilde{\tau}) &=& \gamma^{}_{\tilde{\tau}}\left(\tilde{x}^3 + v^{}_{\tilde{\tau}}\tilde{x}^0 \right) 
		 = \gamma^{}_{\tilde{\tau}}\left(\tilde{x}^3 - v^{}_{\tilde{\tau}}\tilde{r}\right), \label{MintoPOV}
\end{eqnarray}
and express $dx^\mu$ as the linear combinations of $d\tilde{\tau}$, $d\tilde{x}^1$, $d\tilde{x}^2$, and $d\tilde{x}^3$, or more conveniently, with the spatial part in the spherical coordinates $(\tilde{x}^1, \tilde{x}^2, \tilde{x}^3) = (\tilde{r} \sin\tilde{\theta} \cos\tilde{\varphi}, \tilde{r} \sin\tilde{\theta} \sin\tilde{\varphi}, \tilde{r}\cos\tilde{\theta})$. Then, one obtains the line element
\begin{eqnarray}
  ds^2 &=& -\left[ (1 + a\tilde{r}\cos\tilde{\theta})^2 -(a\tilde{r})^2\right] d\tilde{\tau}^2 + 
	2 \left( d\tilde{r}  +  a \tilde{r}^2 \sin \tilde{\theta} d\tilde{\theta} \right) d\tilde{\tau}
	+ \tilde{r}^2\left( d\tilde{\theta}^2 + \sin^2\tilde{\theta} d\tilde{\varphi}^2\right), \label{ObsCoSpher}
\end{eqnarray}
which is almost a special case of those in Refs. \cite{NU63, Ki69} except that here $g^{}_{\tilde{r}\tilde{\tau}}=g^{}_{\tilde{\tau}\tilde{r}}=1$ instead of $-1$ because we are looking at the past light cones (advanced coordinates) for the observer instead of the future light cones (retarded coordinates \cite{Po03}) considered in Refs. \cite{NU63, Ki69}. 
%Here we have $m=0$ since we have implicitly assumed that our localized observer does not disturb the spacetime geometry 
\footnote{The line element (\ref{ObsCoSpher}) is almost identical to Eq.(4) in Ref. \cite{Bo94} except the sign of 
$g^{}_{\tilde{\tau}\tilde{r}}$ ($g^{}_{ur}$ there).}

A coordinate singularity in the observational coordinates (\ref{ObsCoSpher}) occurs at $(1+a\tilde{r}\cos\tilde{\theta})^2-(a\tilde{r})^2 = 0$, which implies $(x^0)^2 - (x^3)^2 =0$ in Minkowski coordinates from (\ref{MintoPOV}). In particular, the hypersurface $V=x^0+x^3=0$ is the past horizon of radar coordinates (\ref{UAradar}) 
for the uniformly accelerated observer, on which 
\begin{equation}
\tilde{r} = \frac{a^{-1}}{1-\cos\tilde{\theta}} \label{pasthor}
\end{equation}
is independent of $\tilde{\tau}$ and the azimuthal angle $\tilde{\varphi}$. 
Eq. (\ref{pasthor}) indicates that in view of the uniformly accelerated observer the past horizon is a static paraboloid of revolution with the focus at $\tilde{r}=0$ (where the observer is located), the semi-latus rectum $a^{-1}$, and the open end in the direction of acceleration (Figure \ref{intersec}). 

A person standing on the surface of the Earth experiences a roughly uniform gravitational acceleration $a =g= 9.8$ ${\rm m/s^2}$. If this gravitational acceleration were strictly uniform in the Universe, that person would see a static paraboloidal past horizon (Figure \ref{intersec} (right)) with the semi-latus rectum $c^2/a \approx 10^{16}$ m $\approx 1$ light-year, which is much larger than the scale of the Earth.

\begin{figure}
\includegraphics[width=3.8cm]{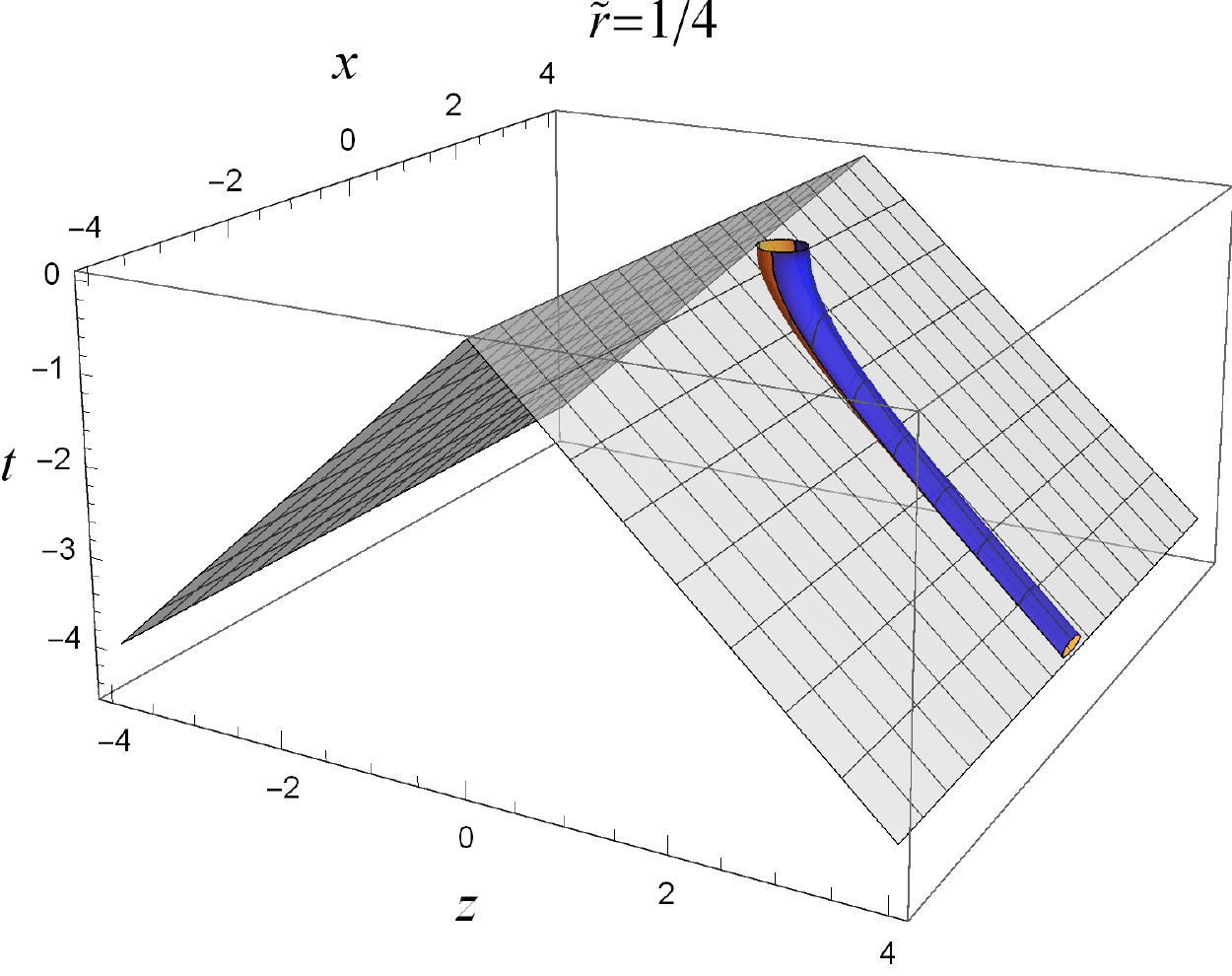} %{Contour_rp25.pdf}
\includegraphics[width=3.8cm]{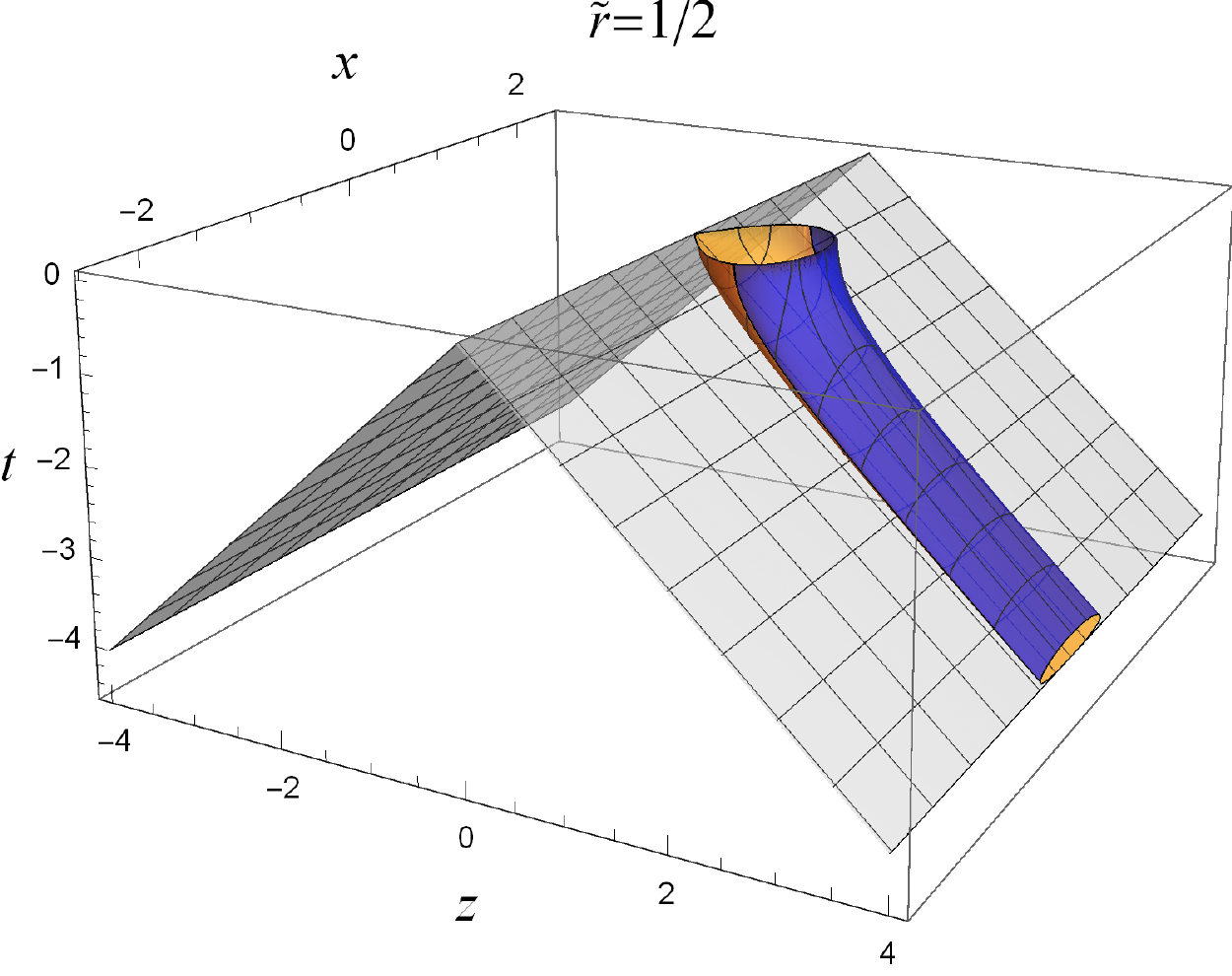} %{Contour_rp5.pdf}
\includegraphics[width=3.8cm]{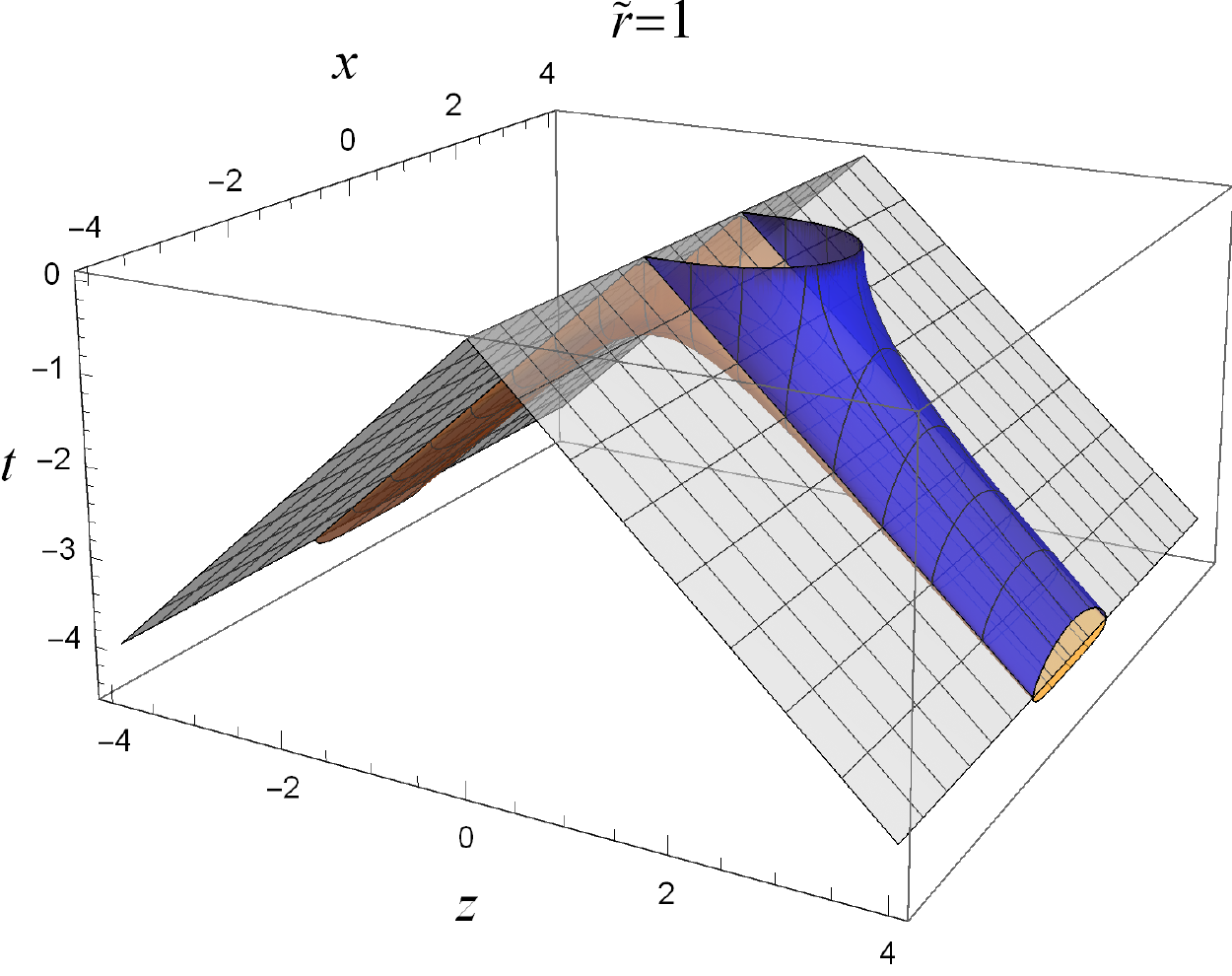} %{Contour_r1.pdf}
\includegraphics[width=5.5cm]{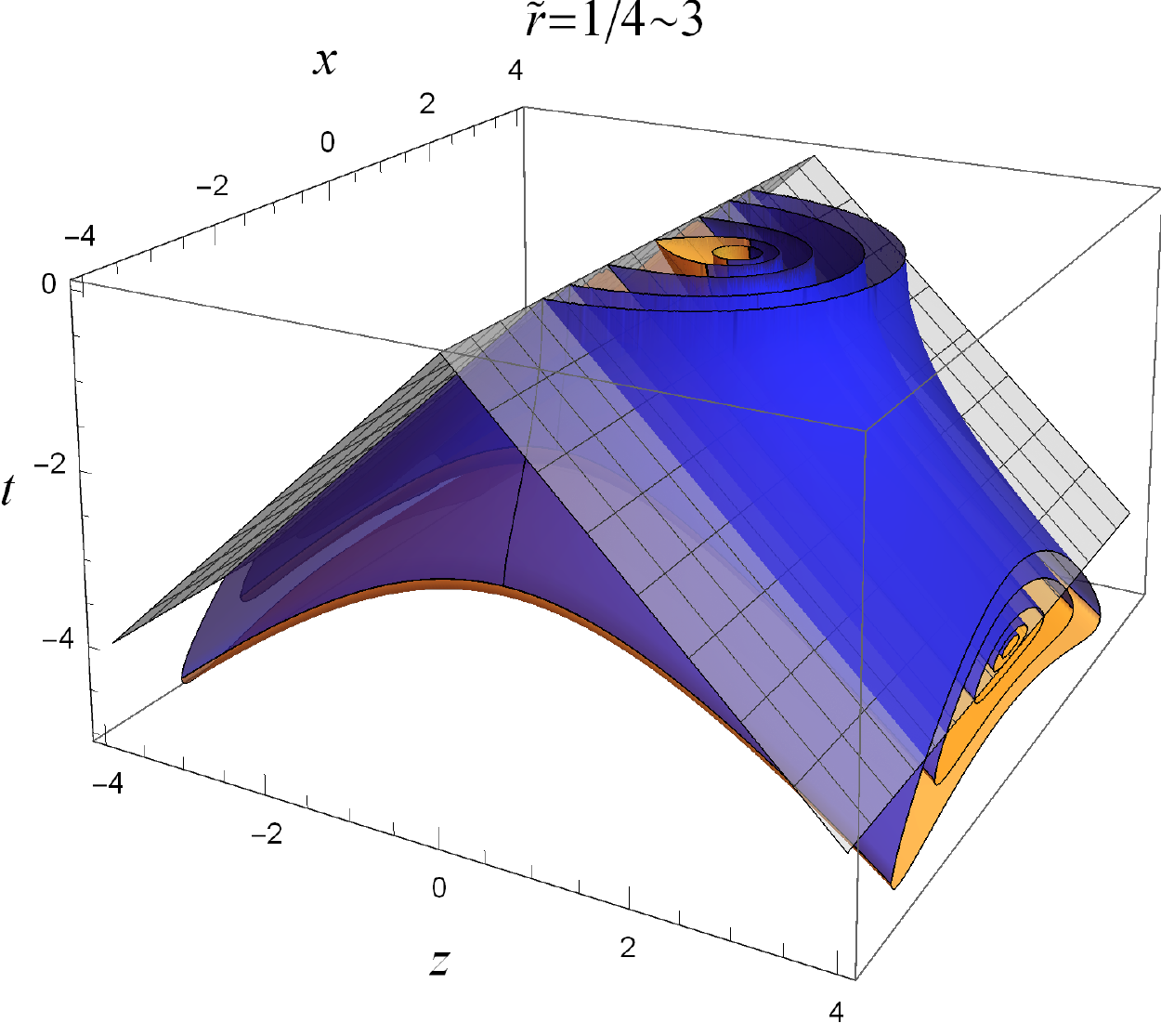} %{Contour_all.pdf}
\caption{Contour surfaces of constant advanced distance $\tilde{r}$ in view of the uniformly accelerated observer moving along $z^\mu(\tau)=(a^{-1}\sinh a\tau, 0,0,a^{-1}\cosh a\tau)$ with $a=1$, represented in Minkowski coordinates with $\varphi$ suppressed. The blue and orange surfaces represent the region with $0\le\tilde{\theta}\le\pi/2$ and $\pi/2<\tilde{\theta}\le\pi$ in the localized observer's point of view ($\tilde{\theta}=0$ and $\pi$ in the $+z$- and $-z$-directions, respectively). One can see the variation of the borders ($\tilde{\theta}=\pi/2$) in plots of different $\tilde{r}$ in Minkowski coordinates. The hypersurfaces $t=z$ and $t=-z$ (gray) are the event and past horizons, respectively, for the uniformly accelerated observer.}
\label{rcontour}
\end{figure}

%Extended to the spacetime with $\rho^2= (x^1)^2 +(x^2)^2 > 0$, 
The constant-$\tilde{r}$ hypersurfaces in Minkowski coordinates are given by
\begin{equation}
  (x^0)^2 - (x^3)^2 = \rho^2 \mp 2a^{-1}\sqrt{\tilde{r}^2- \rho^2}- a^{-2} \label{constrtilde}
\end{equation}
from (\ref{MintoPOV}), where ``$-$" and ``$+$" correspond to the cases with $\tilde{\theta} \in [0, \pi/2]$ and $(\pi/2, \pi]$, respectively.
In Figure \ref{rcontour} we show an example of constant-$\tilde{r}$ hypersurfaces. 
While every constant-$\tilde{r}$ hypersurface with $\tilde{r} < 1/(2a)$ is timelike (Figure \ref{rcontour} (left)), 
the constant-$\tilde{r}$ hypersurfaces with $\tilde{r}\ge 1/(2a)$ are not timelike everywhere in Minkowski coordinates (other plots in Figure \ref{rcontour}). 
In the plane of the observer's motion $x^1=x^2=0$, the line element (\ref{ObsCoSpher}) reduces to
\begin{equation}
  ds^2 	= -(1 \pm 2a\tilde{r}) d\tilde{\tau}^2 + 2d\tilde{r} d\tilde{\tau} \label{RindObs2D}
\end{equation}
with $+$ and $-$ for $\tilde{\theta}=0$ and $\pi$, respectively.
From (\ref{constrtilde}) with $\rho=0$, the contours of $\tilde{r}$ in (\ref{RindObs2D}) are timelike in wedge R ($\tilde{\theta}=0$ and $\tilde{r}>0$,  or $\tilde{\theta}=\pi$ and $2a\tilde{r}<1$) but spacelike in wedge P ($\tilde{\theta}=\pi$ and $2a\tilde{r}>1$) on the $tz$-plane in Minkowski coordinates (Figure \ref{POVMin} (middle)).
Such a pattern is similar to those contours of $\zeta$ in the maximally extended Rindler radar coordinates in wedges R and P in Figure \ref{Rind} (right), though they are not exactly the same.

\begin{figure}
\includegraphics[width=5.5cm]{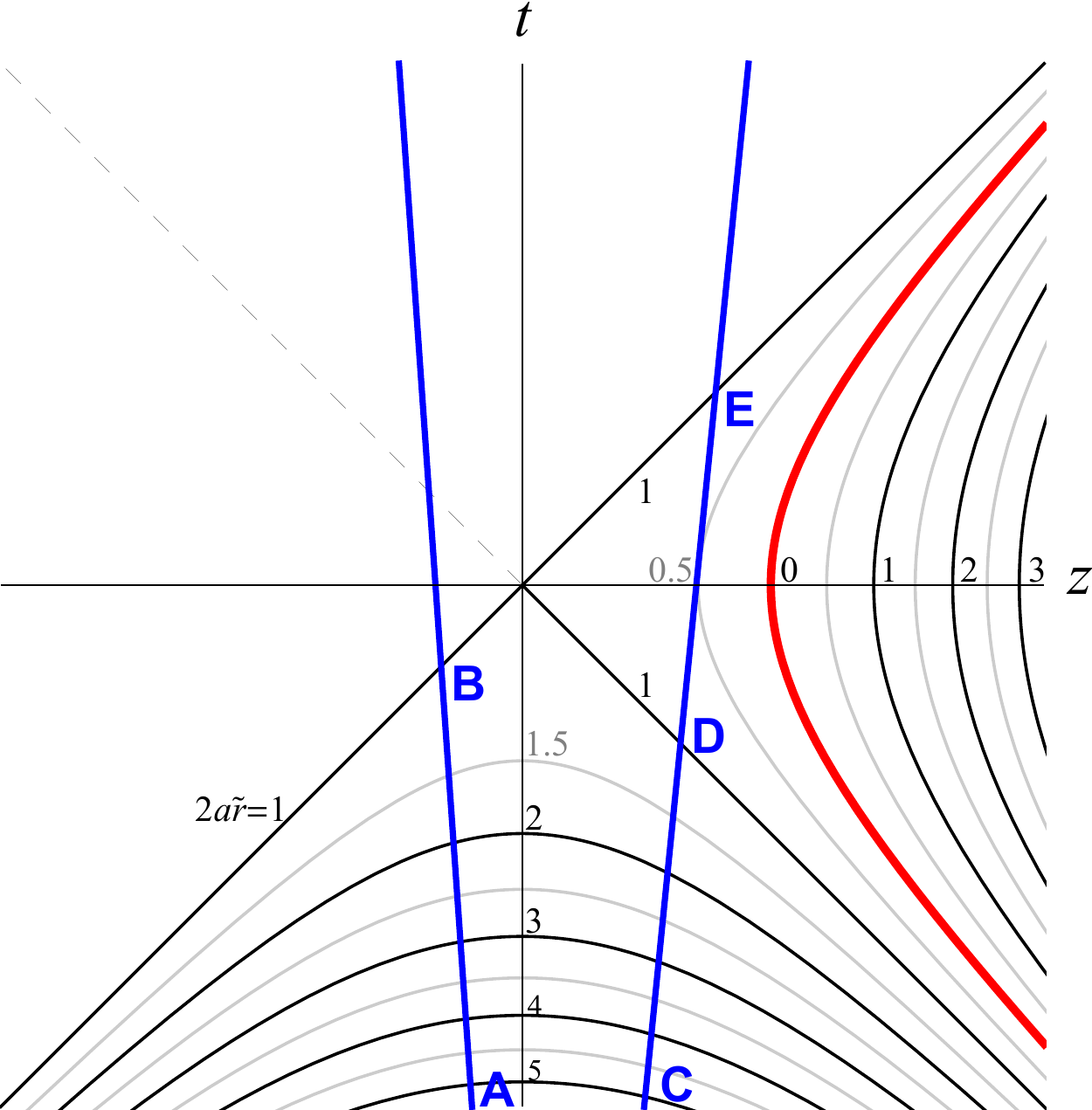} %{UAObs2.pdf}
\caption{Two emitters which are point-like test particles going along the timelike worldlines {\sf AB} and {\sf CDE} (blue) emit light rays continuously to the uniformly accelerated observer at proper acceleration $a$ (red worldline). The observer would see in the direction of $\tilde{\theta}=\pi$ that one emitter started with {\sf A}, went toward the observer and stopped right behind the past horizon ({\sf B}). The other emitter would be seen as started with point {\sf C}, went toward the observer and crossed the past horizon ({\sf D}), reached the minimum distance $\tilde{r}\approx 0.5/(2a)$ of this trip, and then dropped back and eventually stopped around the past horizon and never crossed it again ({\sf E}).}
\label{2WL}
\end{figure}

When a point-like light source moving along a timelike worldline in Minkowski coordinates is seen behind the past horizon (i.e., in wedge P) by the localized observer (e.g. segment {\sf AB} in Figure \ref{2WL}), the source's $\tilde{r}$ and $\tilde{\theta}$ can never both be constants of time in the observer's point of view, a fact associated with the signature change of $g_{\tilde{\tau}\tilde{\tau}}$ in (\ref{ObsCoSpher}) (or $g_{\tilde{t}\tilde{t}}$ in (\ref{ObsCotr})). Thus, the past horizon is also a static limit surface for the uniformly accelerated observer. Moreover, the accelerated observer will see that all the point-like sources %moving along timelike worldlines 
not going to future null infinity will eventually approach the past horizon and then stop there. For example, in Figure \ref{2WL}, two point-like emitters moving along blue timelike worldlines {\sf AB} and {\sf CDE} emit light rays continuously to the uniformly accelerated observer. %going along the red worldline. 
The observer would see in the direction of $\tilde{\theta}=\pi$ that the emitter started with point {\sf A} at distance $\tilde{r}\approx 5/(2a)$ would go toward the observer and then stop right behind the past horizon ({\sf B}) at $\tilde{r}= 1/(2a)$ and never cross it. The other emitter would be seen also in the direction of $\tilde{\theta}=\pi$ and started at the same distance $\tilde{r}\approx 5/(2a)$ ({\sf C}) but at some moment $\tilde{\tau}$ earlier than the time when event {\sf A} is observed. It would go toward the observer and cross the past horizon ({\sf D}), reach the minimum distance $\tilde{r}\approx 0.5/(2a)$ of this trip, and then drop back and eventually stop in front of the past horizon without crossing the past horizon again ({\sf E}).

If a collection of point-like sources are not exactly in the directions $\theta = 0$ or $\pi$ of the localized observer, as they are observed to approach the past horizon, they would also be concentrating toward the direction of the observer's velocity as special relativity implies \cite{HP10}. 

A constant-$\tilde{\tau}$ hypersurface in our observational coordinates (\ref{ObsCoSpher}) is the past light cone of $z^\mu(\tilde{\tau})$ and so $\tilde{\tau}$ is a null coordinate. One may further define a time coordinate $\tilde{t}$ by letting
\begin{equation}
  d\tilde{t} \equiv d\tilde{\tau} - \frac{d\tilde{r} + a \tilde{r}^2\sin\tilde{\theta} d\tilde{\theta}}
	{(1+a \tilde{r}\cos \tilde{\theta})^2 -(a\tilde{r})^2} 
	=  d\tilde{\tau} + \frac{d (\tilde{r}^{-1} + a \cos\tilde{\theta})}
	{(\tilde{r}^{-1} + a \cos\tilde{\theta})^2 -a^2}, \label{tildet}
\end{equation}
and then (\ref{ObsCoSpher}) becomes
\begin{equation}
  ds^2 = -\left[ (1 + a\tilde{r}\cos\tilde{\theta})^2 -(a\tilde{r})^2\right] d\tilde{t}^2 + 
	\frac{( d\tilde{r}+a\tilde{r}^2 \sin \tilde{\theta} d\tilde{\theta} )^2}{(1 + a\tilde{r}\cos\tilde{\theta})^2 -(a\tilde{r})^2}
	+ \tilde{r}^2\left( d\tilde{\theta}^2 + \sin^2\tilde{\theta} d\tilde{\varphi}^2\right), \label{ObsCotr}
\end{equation}
where the constant-$\tilde{t}$ slices are timelike in wedge R and spacelike in wedge P, and all of them will intersect at the origin of Minkowski coordinates as the observer goes to future infinity, similar to the Rindler-time slices in Figure \ref{Rind} (right). 
From (\ref{tildet}), one has
\begin{equation}
     \tilde{t} = \tilde{\tau}+ \frac{1}{2a}\ln \left| 
		\frac{1+a\tilde{r}\cos\tilde{\theta}-a\tilde{r}}{1+a\tilde{r}\cos\tilde{\theta}+a\tilde{r}}\right| 
\end{equation}
which goes to $-\infty$ in both wedges R and P as the observed events goes to the past horizon $x^0+x^3=0$ (cf. Eq.(\ref{pasthor})), and so the past horizon is part of the past infinity with respect to $\tilde{t}$ for the observer. In the plane of the observer's motion, the line element (\ref{ObsCotr}) reduces to $ds^2 = -(1 \pm 2a\tilde{r}) d\tilde{t}^2 + (1 \pm 2a\tilde{r})^{-1}d\tilde{r}^2$ for $\tilde{\theta}=0$ and $\pi$ (cf. (\ref{RindObs2D})).

Interestingly enough, when we fix $\tilde{\theta}=\pi/2$, $\tilde{\varphi}=$constant, and $d\tilde{\theta}=d\tilde{\varphi}=0$, the line element (\ref{ObsCotr}) on this slice looks very much like the static de Sitter coordinates (\ref{deStat}) with the angular dimensions suppressed and $a$ here being identified as the Hubble constant $H$ there. Indeed, the static de Sitter coordinates have some properties similar to (\ref{ObsCotr}), as we will discuss in next section.

\subsection{Non-uniform linear acceleration}
\label{NULA}

As the motion of the observer is switched from non-inertial to inertial, the spacelike part of the constant-$\tilde{r}$ hypersurfaces behind the past horizon for the accelerated localized observer will evolve to timelike surfaces. These constant-$\tilde{r}$ hypersurfaces in our observational coordinates will not be smooth if the observer's acceleration is suddenly changed. 
%while the 4-velocity changes continuously. 
For example, in Figure \ref{POVMin} (right), when the acceleration suddenly drops to zero, while the tangent vector of the worldline of the observer evolves continuously, the constant-$\tilde{r}$ hypersurfaces are not differentiable around the past light cone of the moment that the observer changes acceleration. To make it differentiable the observer's acceleration has to be changed smoothly. The radar coordinates for the same observer behave better: the hypersurfaces of constant radar distance evolve in the same way as the observer's motion; namely, the first derivatives are continuous \cite{DG01}.

\subsection{Spinning observer without center-of-mass motion}
\label{spinO}

Suppose an observer is situated at the origin and spinning about the $z$-axis at a constant angular frequency $\omega$. 
Radar coordinates of the event $E$ at $(t, z, \rho, \varphi)$ in Minkowski cylindrical coordinates can be constructed by the spinning localized observer in a way similar to those for a non-spinning rest observer.
If a radar signal is emitted by the observer at $\tau_i$ and the echo from the event is received at $\tau_f$ in the observer's clock, then  
the radar time and radar distance of the event are again $t=(\tau_f+ \tau_i)/2$ and $r = (\tau_f-\tau_i)/2$ for the observer. The polar angle of the event $\theta = \tan^{-1}(z/\rho)$ is a constant of time for the observer, and so the radar polar angle is still $\theta$. The only modification is that the radar azimuthal angle $\varphi'$ of the event should be determined by the average of the 
%emitting and receiving 
azimuthal angles of arrival and departure of the radar signal in the local frames of the observer at $\tau_f$ and $\tau_i$, respectively: $\varphi' = (\varphi(\tau_f) + \varphi(\tau_i))/2 = \varphi - \omega (\tau_f+\tau_i)/2 = \varphi - \omega t$. Thus, the line element in radar coordinates can be obtained from Minkowski cylindrical coordinates by the transformation $(t', z', \rho', \varphi') = (t, z, \rho, \varphi - \omega t)$,
\begin{equation}
  ds^2 = -(1 -\omega^2 \rho'^2) dt'^2 + 2 \omega \rho'^2 dt' d\varphi' + dz'^2+ d\rho'^2 + \rho'^2 d\varphi'^2 ,
	\label{Borncoord}
\end{equation}
which is the rotating cylindrical coordinates \cite{LL71, PV00, Bo09}. It is well known that
clocks along a closed curve in this coordinate system cannot be synchronized uniquely since $g^{}_{t'\varphi'}\not=0$ \cite{LL71}. 
%Eqs. (88.5), (89.2)
One may define a new time coordinate as
\begin{equation}
  dT' = dt' - \frac{\omega \rho'^2 d\varphi}{1-\omega^2 \rho'^2}
\end{equation}
to diagonalize (\ref{Borncoord}) into
\begin{equation}
  ds^2 = -\left(1 -\omega^2 \rho'^2\right) dT'^2+ \frac{\rho'^2 d\varphi'^2}{1-\omega^2\rho'^2} + dz'^2+ d\rho'^2,
	\label{BornNewT}
\end{equation}
where the timelike and spacelike properties of $T'$ and $\varphi'$ coordinates will be switched when the observed events are crossing the cylinder $\rho'=1/\omega$. 

The same event $E$, now represented as $(t,r, \theta, \varphi)$ in Minkowski spherical coordinates, will be observed by the spinning observer 
at her proper time $\tilde{\tau}=t+r$. At that moment the observer will see the event in the direction $(\tilde{\theta},\tilde{\varphi}) = (\theta, \varphi - \omega\tilde{\tau}) =(\theta, \varphi - \omega (t+r))$ at the distance $\tilde{r}=r$ away from the observer.
Thus the observational coordinates for the spinning observer read 
\begin{eqnarray}
  ds^2 &=& 
	-d\tilde{\tau}^2 + 2d\tilde{\tau}d\tilde{r} + \tilde{r}^2 \left(d\tilde{\theta}^2 + 
	\sin^2\tilde{\theta} (d\tilde{\varphi}+\omega d\tilde{\tau})^2\right)	\nonumber\\
	&=& -\left(1-\omega^2\tilde{r}^2 \sin^2\tilde{\theta}\right)d\tilde{\tau}^2 +2d\tilde{\tau}
	\left( d\tilde{r}+ \omega \tilde{r}^2 \sin^2\tilde{\theta} d\tilde{\varphi} \right) + \tilde{r}^2 \left(d\tilde{\theta}^2 +
	\sin^2\tilde{\theta} d\tilde{\varphi}^2 \right). \label{POVspin}
\end{eqnarray}
Define a new time coordinate $\tilde{t}$ as
\begin{equation}
  d\tilde{t} = d\tilde{\tau} -\frac{d\tilde{r}+\omega\tilde{r}^2\sin^2\tilde{\theta}d\tilde{\varphi}}{1-\omega^2\tilde{r}^2\sin^2 \tilde{\theta}};
\end{equation} 
one can rewrite (\ref{POVspin}) as
\begin{equation}
  ds^2 = -\left(1-\omega^2\tilde{r}^2 \sin^2 \tilde{\theta}\right)d\tilde{t}^2 + 
	\frac{\left(d\tilde{r}+\omega\tilde{r}^2 \sin^2 \tilde{\theta}d\tilde{\varphi}\right)^2}{1-\omega^2\tilde{r}^2 \sin^2 \tilde{\theta}}
	+\tilde{r}^2 \left( d\tilde{\theta}^2 + \sin^2\tilde{\theta}d\tilde{\varphi}^2\right).
	\label{POVspint}
\end{equation}
%there is no coordinate singularity at $\tilde{\theta}=0$ and $\pi$.
Transformed to the cylindrical coordinates, $\tilde{\rho} = \tilde{r} \sin \tilde{\theta}$ and $\tilde{z} = \tilde{r} \cos \tilde{\theta}$,
the above line element becomes
\begin{equation}
  ds^2 = -\left(1-\omega^2\tilde{\rho}^2\right)d\tilde{t}^2 + \frac{\tilde{\rho}^2}{1-\omega^2\tilde{\rho}^2}\left[d\tilde{\varphi} + 
	\frac{\omega(\tilde{\rho} d\tilde{\rho} +\tilde{z} d\tilde{z})}{\sqrt{\tilde{\rho}^2+\tilde{z}^2}}\right]^2+d\tilde{z}^2+d\tilde{\rho}^2, 
	\label{POVspinCyl}
\end{equation}
which is not the same as (\ref{BornNewT}) because here $\tilde{\varphi} = \varphi -\omega(t+r)$ in observational coordinates but there $\varphi' =\varphi -\omega t$ in radar coordinates.

The coordinate-singularity cylinder of perceived radius $\tilde{\rho} =1/\omega$ is a static limit surface, beyond which nothing along a timelike worldline in Minkowski coordinates can be at rest in view of the spinning observer. Indeed, for $\tilde{\rho} > 1/\omega$, $d\tilde{t}$ becomes spacelike and $d\tilde{r}+\omega \tilde{r}^2 \sin^2\tilde{\theta} d\tilde{\varphi}$ becomes timelike. For our Earth, $\omega \approx 2\pi/(86164 \, {\rm s})=7.29 \times 10^{-5} {\rm s}^{-1}$, %$2\pi/\omega =1$ day $\approx 86400$ seconds, 
so the static limit surface would be positioned at $\tilde{\rho} \approx 4.11 \times 10^{12}$m ($\approx 1.37\times 10^4$ light-seconds) away from the Earth.

Two features of the observational and radar coordinates here are different from those for a non-spinning, uniformly accelerated observer. First, radar coordinates (\ref{BornNewT}) have a nontrivial coordinate singularity at $\rho' = 1/\omega$, which is the same static limit surface in the observational coordinates (\ref{POVspinCyl}) for the same observer, while the radar coordinates for a uniformly accelerated observer (\ref{UAradar}) are regular for all finite values of the coordinates.
Second, assuming the emitting and receiving operations of the localized observer have been started early around past timelike infinity,
then radar coordinates (\ref{Borncoord}) and observational coordinates (\ref{POVspint}) will cover almost the same region in the Penrose diagram of Minkowski space except the neighborhood of past null infinity. 
In contrast, for a non-spinning uniformly accelerated observer that started the operations around past null infinity, the spacetime region covered by her observational coordinates are clearly larger than the region covered by her radar coordinates. 

%%%%%%%%%%%%%%%%%%%%%%%%%%%%%%%%%%%%%%%%%%%%%%
\section{Comoving observer in De Sitter Space}
\label{StatOdeS}

Similar to Rindler coordinates in Minkowski space, the static and flat de Sitter coordinates do not cover the whole de Sitter space.
%One may wonder if the observational and radar coordinates of a localized observer in de Sitter space can be related to the static, flat, or other conventional de Sitter coordinates. 
Below, we are constructing the observational and radar coordinates for a comoving observer localized in de Sitter space to see if they can be 
related to the static, flat, or other conventional de Sitter coordinates. .
%Then we will see if they have the features similar to those in Minkowski space.

\subsection{Observer localized at the origin}
\label{deSOO}

\begin{figure}
\includegraphics[width=6.5cm]{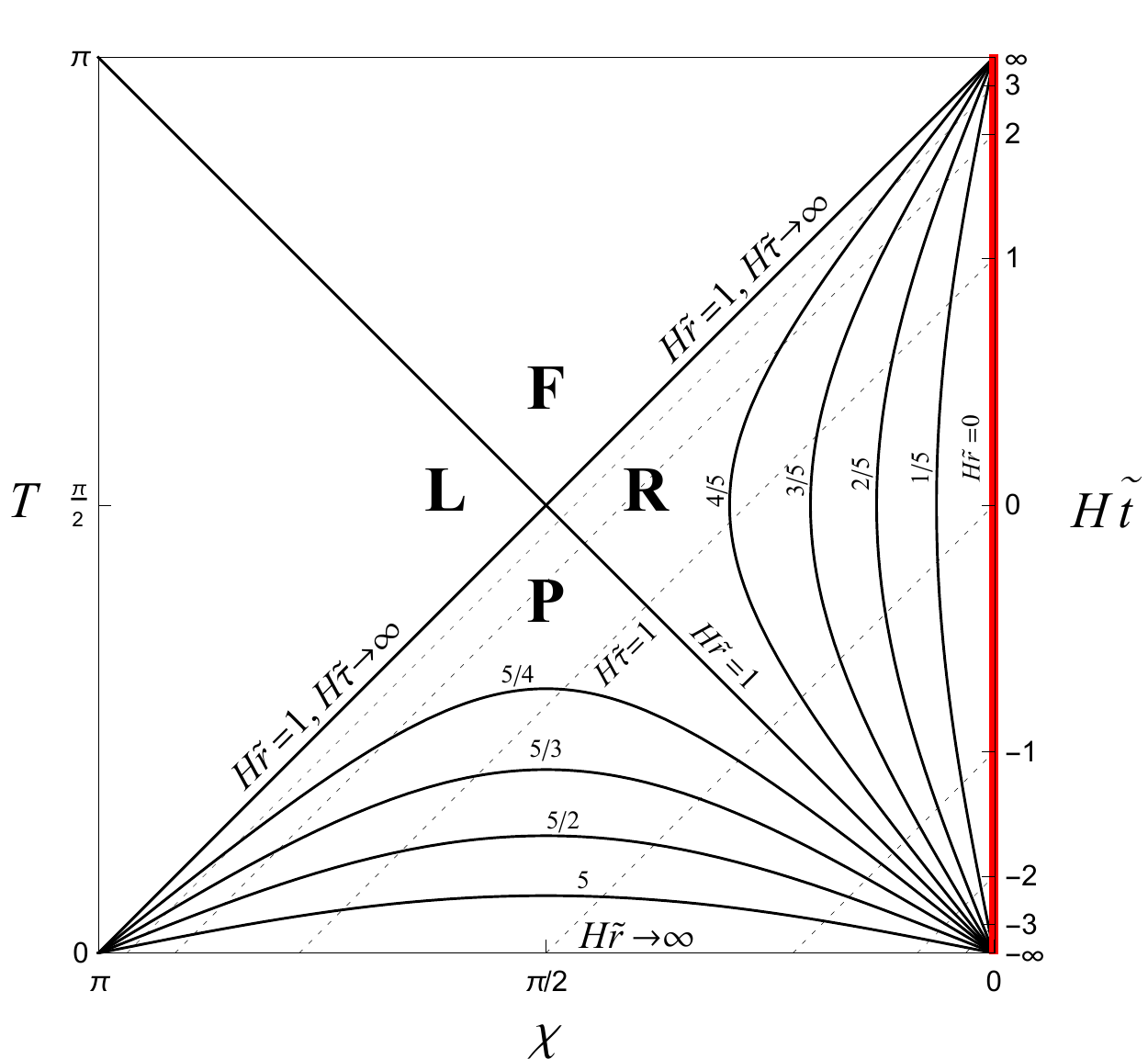} %{POVdeS4.pdf}
\includegraphics[width=6.5cm]{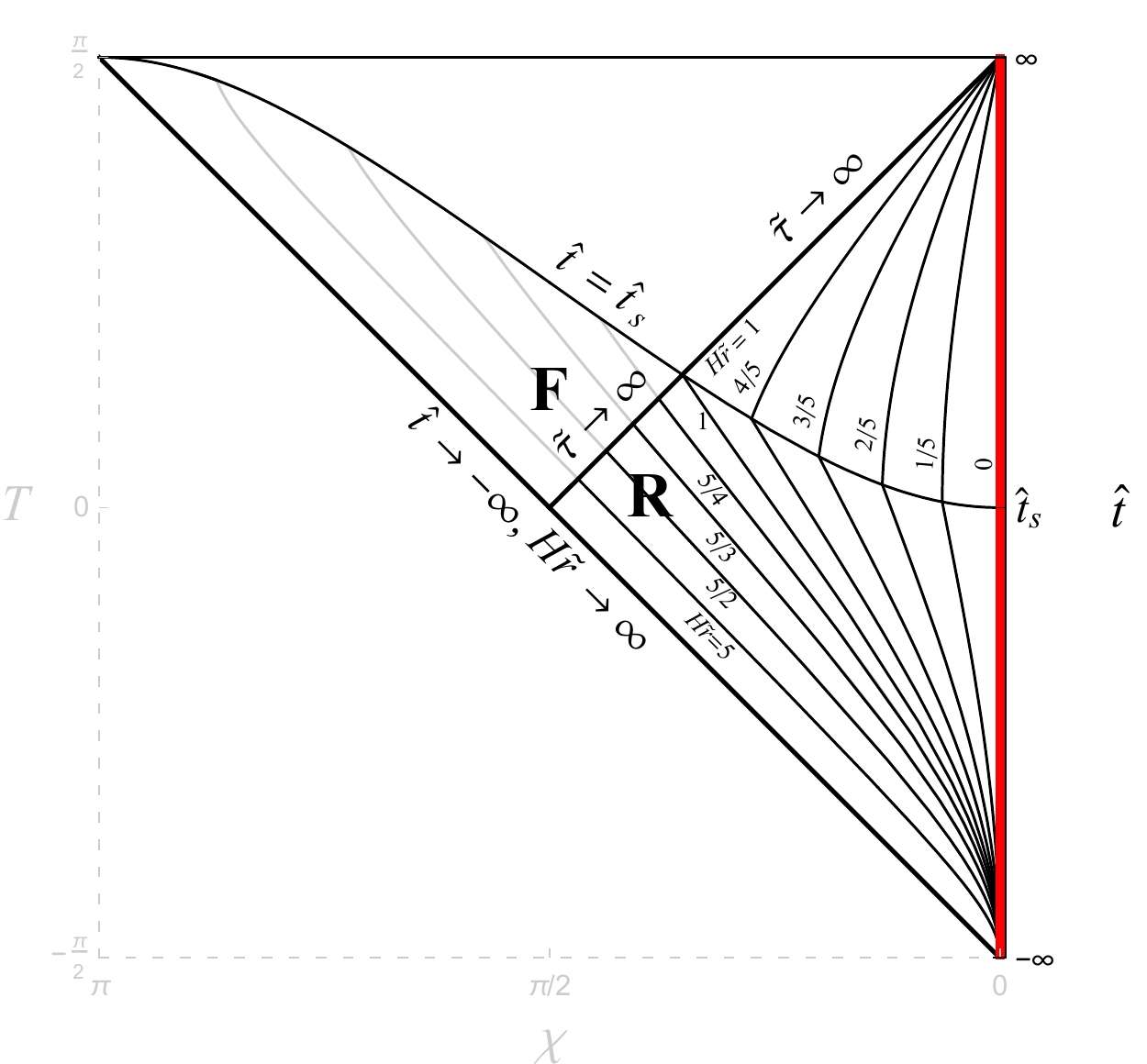} %{POVnEI2.pdf}
\caption{(Left) The Penrose diagram for de Sitter space, with observational coordinates for an observer at $\tilde{r}=\chi=0$ (red). The solid curves represent the constant $\tilde{r}$ (affine distance from the localized observer) hypersurfaces, and the dotted lines represent the constant $\tilde{\tau}$ (proper time of the localized observer) hypersurfaces. The hypersurfaces $T= \chi$ and $T=\pi-\chi$ are the past horizon and the event horizon for the localized observer, respectively. $H\tilde{r}=1$ on both horizons. (Right) The non-eternal inflationary universe considered in section \ref{NEInflat}. Above and below the hypersurface $\hat{t}=\hat{t}_s$ are pieces of de Sitter and Minkowski spaces, respectively. There is no need of region L or P here since the region F $\cup$ R is geodesically complete. Both the observational and radar coordinates for the observer localized at $\rho = \chi =0$ cover region R only.}
\label{POVdeS1}
\end{figure}

Consider the global coordinates in de Sitter space \cite{HE73, Mo85}, 
\begin{eqnarray}
  ds^2 &=& - dt^2 + H^{-2}\cosh^2 H t\left[ d\chi^2 + \sin^2 \chi d\Omega_{\rm II} \right], \label{deSK1}\\
	&=& (H \sin T)^{-2} \left[ -dT^2 + d\chi^2+\sin^2\chi d\Omega_{\rm II}\right]
\end{eqnarray}
where $d\Omega_{\rm II} = d\theta^2 + \sin^2\theta d\varphi^2$, $H$ is the Hubble constant, and $T = 2\tan^{-1}[\tanh (Ht/2)]$  %e^{Ht}$.
so that $T=0$ when $t=0$. 
The Penrose diagram of the de Sitter space with $\theta$ and $\varphi$ suppressed can thus be represented as a square with
$T \in (-\pi/2, \pi/2)$ and $\chi \in (0, \pi)$ in the $T\chi$-plane, as shown in Figure \ref{POVdeS1} (left).
Let 
\begin{eqnarray}
  H\tilde{r} &=& \cosh Ht \, \sin \chi = \frac{\sin\chi}{\sin T}, \label{roft1X}\\
  H\tilde{t} &=& \tanh^{-1}\left( \tanh Ht\, \sec\chi \right).
\end{eqnarray}
Then we get 
%140517, 140522
the static de Sitter coordinates
\begin{equation}
  ds^2 = -(1-H^2 \tilde{r}^2)d\tilde{t}^2 + \frac{d\tilde{r}^2}{1-H^2 \tilde{r}^2} + \tilde{r}^2 d\Omega_{\rm II},
	\label{deStat}
\end{equation} 
where a coordinate singularity occurs at $\tilde{r}=1/H$.
For an observer situated at $\tilde{r}=\chi=0$, the coordinate time $\tilde{t}$ in (\ref{deStat}) is identical in value to the coordinate time $t$ in (\ref{deSK1}) as well as the localized observer's proper time. Interestingly enough,
$\tilde{r}$ here is actually the angular diameter distance %Eq. (14.4.17) in \cite{We72}
and the affine distance of the null geodesic from an event $(\tilde{t},\tilde{r},\theta, \varphi)$ 
to the observer at the spatial origin $\chi=\tilde{r}=0$, as will be shown in section \ref{APflatdeS}. 
For $\tilde{r} < 1/H$, the static coordinates (\ref{deStat}) have the metric component $g_{\tilde{t}\tilde{t}}>0$ and cover the region R 
in Figure \ref{POVdeS1} (left), which is the counterpart of wedge R in Minkowski space.
Outside region R, one can keep using $\tilde{r}$ in (\ref{roft1X}), which is well defined 
and ranges from $1/H$ (the past horizon $T=\chi$) to $\infty$ (past null infinity $T=0$) in region P.
Note that from (\ref{roft1X}) one has the contours of $\tilde{r}$ in the $T\chi$-plane as
$\chi = \sin^{-1}[H\tilde{r}\sin T ]$ for $H\tilde{r} < 1$, which are timelike in region R, and 
$T = \sin^{-1}[(H\tilde{r})^{-1}\sin\chi]$ for $H\tilde{r} > 1$, which are spacelike in region P.
The boundaries of the regions, $T=\chi$ and $T=\pi - \chi$ for $H\tilde{r}=1$, are lightlike.

One may define the null coordinate
\begin{equation}
  d\tilde{\tau} = d\tilde{t} + \frac{d\tilde{r}}{1-H^2 \tilde{r}^2};
\end{equation}
then, (\ref{deStat}) becomes
\begin{equation}
   ds^2 = -(1-H^2 \tilde{r}^2)d\tilde{\tau}^2 + 2 d\tilde{r} d\tilde{\tau} + \tilde{r}^2 d\Omega_{\rm II},
	\label{deStatEF}
\end{equation}
which would be a good observational coordinate system to specify the observed events in regions P and R for the observer localized at the origin. Here, $\tilde{\tau}$ is the clock reading of the observer localized at $\tilde{r}=0$ (where $\tilde{\tau} = \tilde{t}$).

%140522-3
A radar coordinate system for the same localized observer can be obtained from (\ref{deStat}) after identifying radar distance
\begin{equation}
  r \equiv H^{-1} \tanh^{-1} H \tilde{r},
\end{equation}
such that $r\to \infty$ as $\tilde{r}\to 1/H$, and
\begin{equation}
  ds^2 = \frac{1}{\cosh^2 H r}\left[ -d\tilde{t}^2 + dr^2 +  \left(\frac{\sinh Hr}{H}\right)^2 
	d\Omega_{\rm II}\right], \label{deSradar}
\end{equation}
which turns out to be conformally equivalent to (\ref{UAradar}) for a uniformly accelerated observer at proper acceleration $a=H$ in Minkowski space. Similar to the Rindler coordinates in Minkowski space, the above radar coordinate system has no nontrivial coordinate singularity for all finite values of the coordinates, and it only covers region R of the de Sitter space (\ref{deSK1}) in the Penrose diagram. Again, the visible universe for the localized observer situated at $\tilde{r}=r=0$ is not restricted in region R where radar coordinates can be defined. The observer can see the events in region P behind the past horizon of radar coordinates (\ref{deSradar}) ($r\to \infty$, or $\tilde{r}=1/H$) and coordinatize those events in observational coordinates (\ref{deStatEF}). 
Moreover, the observer will feel that all the timelike worldlines not going to future infinity of $\chi=0$ will be going toward and eventually stop around the past horizon, which is a sphere of radius $1/H$ centered at the observer.

\subsection{Flat coordinates}
\label{APflatdeS}

Suppose the same comoving observer happens to use the flat de Sitter coordinates, 
\begin{equation}
  ds^2 = -d\hat{t}^2 + a^2(\hat{t}\hspace{.5mm})\left[ d\rho^2 + \rho^2 d\Omega_{\rm II}\right],	\label{deSflat}
\end{equation}
where $a(\hat{t}\hspace{.5mm}) = e^{H\hat{t}}/H$, and the observer is localized at $\rho=0$. Here, $\hat{t}$ and $\rho$ are transformed from the global coordinates (\ref{deSK1}) by \cite{Mo85}
\begin{eqnarray}
  e^{H\hat{t}} &=& \sinh Ht + \cosh H t \cos\chi, \label{toft1X}\\
	\rho &=& \frac{\sin\chi}{\tanh H t + \cos\chi}, \label{rhooft1X}
\end{eqnarray}
and the flat de Sitter coordinates (\ref{deSflat}) cover the region F $\cup$ R bounded by $\chi=0$, $T=\pi/2$, and $\chi=T$ in the Penrose diagram (Figure \ref{POVdeS1}). For every null geodesic from past infinity to the observer at $\hat{t}=\hat{t}_o$, (\ref{deSflat}) implies
\begin{equation}
  \frac{d\hat{t}}{d\rho} = -a\left(\hat{t}(\rho)\right) \label{flatDSnull}
\end{equation}
where the minus sign corresponds to the past light cone. Requiring $\rho=0$ at $t=t_o$, the solution for (\ref{flatDSnull}) is
\begin{equation}
  \rho = e^{-H\hat{t}} - e^{-H\hat{t}_o} \label{rhooft}
\end{equation} 
along the null geodesic. Suppose the observer uses $\lambda \equiv a(\hat{t}_o)\rho$ to parametrize the null geodesic so that the value of 
$\lambda$ matches radar distance in the neighborhood of the localized observer. 
By virtue of the spherical symmetry, the null geodesic can be described by the equation
\begin{equation}
  \frac{d^2 z^\mu}{d\lambda^2} + \Gamma^\mu_{\alpha\beta}\frac{d z^\alpha}{d\lambda}\frac{d z^\beta}{d\lambda} = 
	\kappa(\lambda) \frac{d z^\mu}{d\lambda},
\end{equation}
where we find $\kappa(\lambda) = -2He^{H(\hat{t}(\lambda)-\hat{t}_o)} = -2H/(H\lambda+1)$ after taking $z^{\mu} = (\hat{t}(\lambda), \rho(\lambda), \theta, \varphi)$ and introducing (\ref{flatDSnull}) and (\ref{rhooft}). While $\kappa$ is not zero and so $\lambda$ is not an affine parameter, one can generate an affine parameter $\lambda^*$ from $\lambda$ by solving \cite{Po04}
\begin{equation}
  \frac{d\lambda^*}{d\lambda} = \exp \int^\lambda \kappa(\lambda')d\lambda' = \frac{1}{(H\lambda+1)^2},
\end{equation}
which gives 
\begin{equation}
  \lambda^* = \frac{1}{H}\left(1 - \frac{1}{H\lambda+1}\right) = \frac{1}{H}\left(1 - e^{H(\hat{t}-\hat{t}_o)}\right). \label{affpara}
\end{equation}
with the condition $\lambda^*=0$ at $\lambda=0$.
It turns out that $\lambda^* = a(\hat{t})\rho = H^{-1}\cosh H t \, \sin\chi = \tilde{r}$ from (\ref{affpara}), (\ref{rhooft}), (\ref{toft1X}), (\ref{rhooft1X}), and (\ref{roft1X}); namely, $\lambda^*$ coincides with the radial coordinate $\tilde{r}$, which is nothing but the angular diameter distance $a(\hat{t})\rho$, in the static coordinates (\ref{deStat}) or (\ref{deStatEF}) along the null geodesics on the past light cones of the localized observer. Thus, the coordinates in (\ref{deStatEF}) would be natural observational coordinates after identifying $\tilde{\tau}=\hat{t}_o$ and $\lambda^* = \tilde{r}$. The observer can see through the past horizon $\hat{t}\to -\infty$ of the flat coordinates into region P of de Sitter space, though the flat coordinates do not cover that region.

\subsection{Non-Eternal Inflation} 
\label{NEInflat}

The above result is valid in de Sitter space, corresponding to an eternally inflationary universe. In the usual non-eternal inflation model, a comoving localized observer could not see beyond the null surface $T=\chi$ and the observational coordinates for that observer would cover only wedge R in the $T\chi$-plane, as shown in Figure \ref{POVdeS1} (right). 

For example, suppose an inflation era of a flat spacetime is started with a Minkowski space at $\hat{t}=\hat{t}_s$, and the metric of the spacetime is given by
\begin{equation}
  ds^2 = -d\hat{t}^2 + a^2(\hat{t}\hspace{.5mm})\left( d\rho^2 + \rho^2 d\Omega_{\rm II}\right) 
\label{flatdeSMin}
\end{equation}
where $a(\hat{t}\hspace{.5mm}) = e^{H \hat{t}}/H$ is exponentially growing in time 
for $\hat{t}\ge \hat{t}_s$ and $a(\hat{t}\hspace{.5mm}) = e^{H \hat{t}_s}/H$ is a constant for $\hat{t}< \hat{t}_s$.
Using the inverse transformations of (\ref{toft1X}) and (\ref{rhooft1X}) from $(\hat{t}, \rho)$ to $(t, \chi)$, one can see that the flat coordinates in (\ref{flatdeSMin}) still cover region F $\cup$ R in the Penrose diagram for de Sitter space in the $T\chi$-plane.

For a localized observer at the spatial origin ($\rho=\chi=0$) with the metric in (\ref{flatdeSMin}), the light pulse from the event at $(\hat{t}, \rho, \theta, \varphi)$ is received by the observer at her proper time $\hat{t}_o$ determined by (\ref{rhooft}) for $\hat{t} \ge \hat{t}_s$ and by $\rho = e^{-H\hat{t}_s}- e^{-H \hat{t}_o} +e^{-H\hat{t}_s}(\hat{t}_s - \hat{t})$ for $\hat{t}< \hat{t}_s$.
Thus, the angular diameter distance of the event for the localized observer is 
\begin{equation}
  %d_{ad} = 
	a(\hat{t}\hspace{.5mm})\rho = \left\{\begin{array}{lll}
           H^{-1} \left(1- e^{H (\hat{t} - \hat{t}_o)}\right) & {\rm for} & \hat{t} \ge \hat{t}_s, \\
           H^{-1} \left(1- e^{H (\hat{t}_s - \hat{t}_o)}+\hat{t}_s - \hat{t}\right) & {\rm for} & \hat{t}< \hat{t}_s,
       	 \end{array}\right.
\end{equation}
which diverges as $\hat{t}\to -\infty$.
In other words, the distance from the hypersurface $\hat{t}\to -\infty$ %from the localized observer  
is infinity for the observer at the origin, and the region F $\cup$ R is geodesically complete. In Figure \ref{POVdeS1} (right) there would be nothing behind past infinity %(rather than the past horizon) 
at $T=\chi$ to be visible for the observer.

In Figure \ref{POVdeS1} (right), one can also see that the physical objects with timelike worldlines passing through the event horizon (the null surface labeled $\tilde{\tau}\to\infty$) after the onset of inflation ($\hat{t}>\hat{t}_s$) would appear to go away from the observer and approach the illusory horizon at $\tilde{r}=1/H$ at late times for the observer. The other physical objects, which pass through the event horizon before $\hat{t}_s$, would be observed at late times as frozen at some distances greater than $1/H$, and their clock readings would never reach $\hat{t}_s$, if all the clocks have been synchronized initially at $\hat{t}\to -\infty$.

%%%%%%%%%%%%%%%%%%%%%%%%%%%%%%%%%%%%%%%%%%%%%%%%%%%%%%%%%%%
\section{Observer outside a spherical shell in (1+1)D}
\label{Schwarz2D}

The null geodesics around a black hole in (3+1)D can be complicated even in the simplest case of Schwarzschild spacetime \cite{Pe04, HP10, Lin19b}, and so observational and radar coordinates in terms of perceived and radar distances for a localized observer may not be convenient for analysis. Nevertheless, the observational and radar coordinates for a localized observer in (1+1)D Schwarzschild geometry can be simple enough for us %%%5/4
to gain insights. 

Kruskal coordinates for a (1+1)D Schwarzschild black hole look very similar to Rindler coordinates in (1+1)D Minkowski space.
One may be tempted to think that an observer localized outside an eternal black hole or a collapsing star 
would be able to coordinatize the events behind the past horizon at the Schwarzschild radius in Kruskal coordinates, and would observe that most of the timelike worldlines would eventually approach the past horizon with increasing redshift. 
Similar to the cases in de Sitter space, such a speculation would be true only in the maximally extended Schwarzschild coordinates for an eternal black hole (with the white hole singularity visible by a localized observer outside), but not in the case of a spherical collapsing star, which forms a black hole at late times as in the example below. %for a localized observer always outside.

Consider a (1+1)D spacetime in the presence of a spherical thin shell of mass $M$ and radius $r=r_s > 2M$ \cite{Is66, Po04},
% Eq.(3.75) in Po04: continuity of intrinsic metric on the shell
\begin{equation}
  ds^2 = -A(r) dt^2 + B(r) dr^2,	\label{SchwarzShell}
\end{equation}
where
\begin{eqnarray}
      A(r) = 1/B(r) = 1-\frac{2M}{r} \hspace{.5cm} &{\rm for}&\; r>r_s,\\
			A(r) = A_s \equiv 1-\frac{2M}{r_s}, \hspace{.5cm} B(r)=1 \hspace{.5cm} &{\rm for}&\; r\le r_s,
\end{eqnarray}
and a localized observer outside of the shell is fixed at a constant radial distance $r=r_o> r_s$ from the center of the spherical shell in the above bookkeeper coordinates \cite{TW00}. The event at $(t, r)$ can be specified by radar time 
$t'=(\tau_f+\tau_i)/2 = \sqrt{A_o}\, t$ with $A_o\equiv A(r_o) = 1-(2M/r_o)$, and radar distance 
\begin{eqnarray}
   r' &=& \sqrt{A_o} {\Delta t\over 2} = 
	\sqrt{A_o}\left| \int_{r^{}_o}^r\sqrt{\frac{B(\bar{r})}{A(\bar{r})}}d\bar{r}\right| \nonumber\\
	&=& \left\{\begin{array}{lll}
	      \sqrt{A_o}\left| r-r^{}_o +2M\ln\frac{r-2M}{r^{}_o-2M}\right|  & {\rm for} & r > r_s, \\
			  \sqrt{A_o}\left(\frac{1}{\sqrt{A_s}}(r^{}_s-r)-r^{}_s+r^{}_o-2M\ln\frac{r^{}_s-2M}{r^{}_o-2M}\right)
			& {\rm for} & r \le r_s. \end{array}\right.
	\label{Schwaradar}
\end{eqnarray}
Then (\ref{SchwarzShell}) can be transformed to radar coordinates %181014-3
\begin{equation}
  ds^2 =  \frac{A(r)}{A_o} \left(-dt'^2 +dr'^2\right)		\label{SchwarzRadar}
\end{equation}
for the localized observer. For $r\le r^{}_s$, $g^{}_{r'r'}(=-g^{}_{t't'})$ goes to zero and radar distance $r'(r)$ diverges to infinity as $r^{}_s\to 2M$.

\begin{figure}
\includegraphics[width=6.5cm]{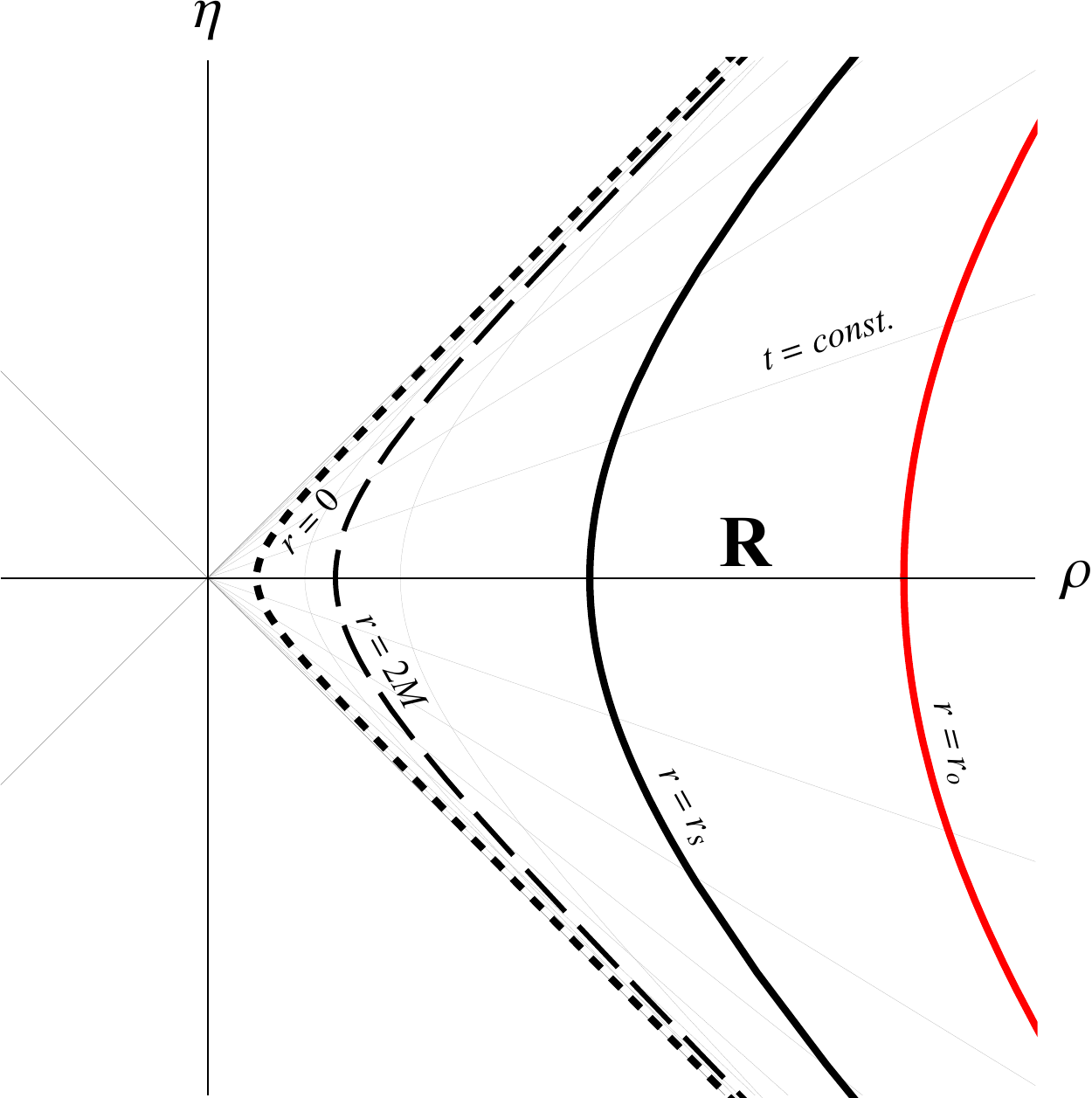} %{Shell_2.pdf} 
\hspace{.5cm}
\includegraphics[width=6.5cm]{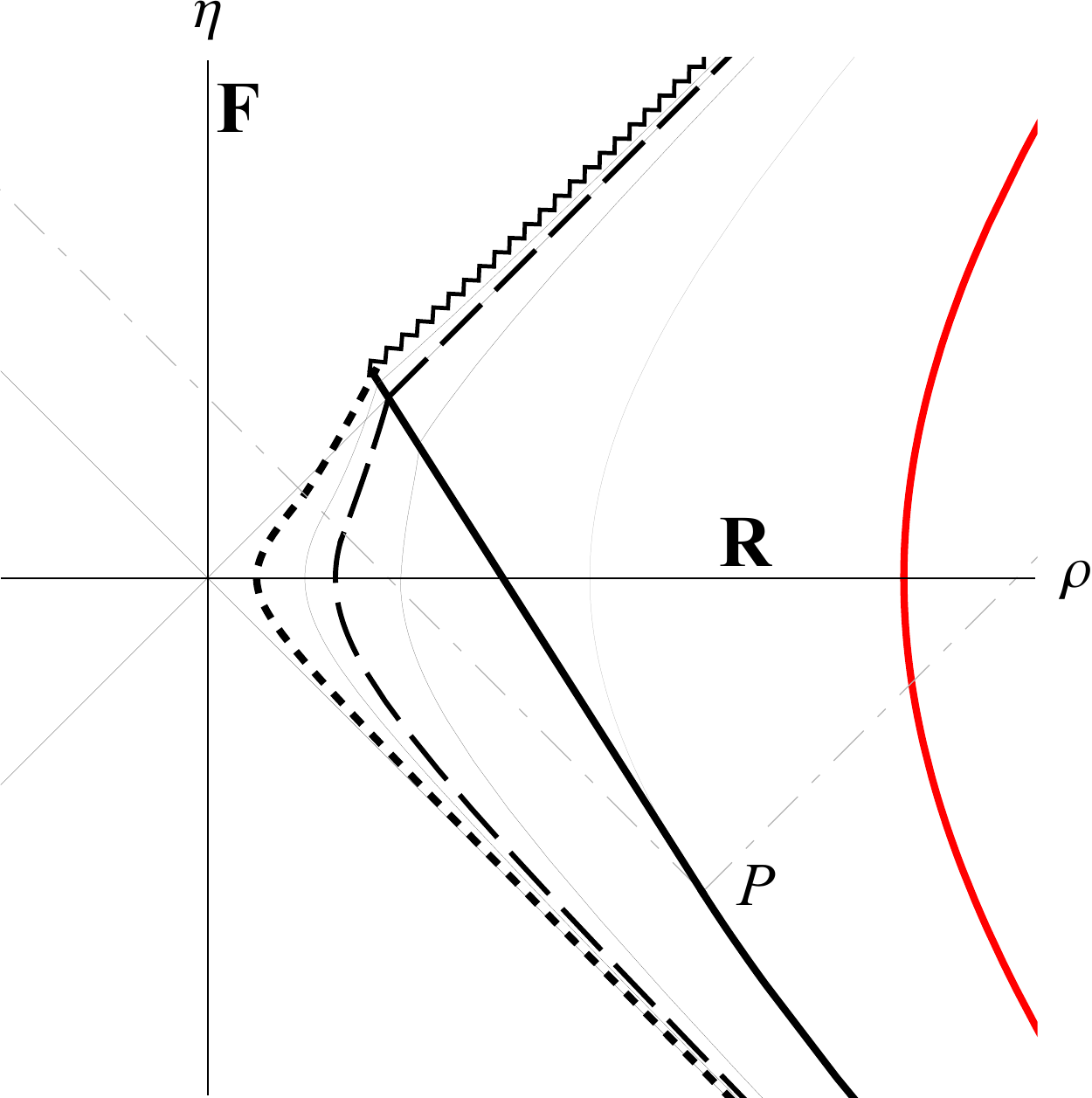} %{Shell_BH2.pdf}
\caption{(Left) Bookkeeper coordinates (\ref{SchwarzShell}) for a spherical massive shell are represented on the $\eta\rho$-plane of the Kruskal-like coordinates (\ref{KruShell}). The hypersurfaces $r=2M$ (dashed) and $r=0$ (dotted) behind the spherical shell $r=r^{}_s$ (thick black) are all in the same ``wedge R." (Right) A speculative scenario of a collapsing shell forming a black hole 
%(not obtained by real calculation; 
(we assume the collapse is adiabatic and dynamics of the metric is ignored). The spacelike zigzag curve represents the singularity at $r=0$ that the shell collapsed into. The dot-dashed lines represent the future light cone emitted from the event $P$ that the shell started to collapse from a constant radius $r=r_s$. Outside the future light cone, the metric is identical to the one in the left plot.}
\label{SchwS}
\end{figure}

Following the same transformations from Schwarzschild coordinates to Kruskal coordinates, (\ref{SchwarzShell}) in the presence of the spherical shell can be transformed into the Kruskal-like coordinates 
\begin{equation}
  ds^2 = \left\{\begin{array}{lll}
	        \frac{16M^2}{r} \exp\left\{-\frac{r}{2M}\right\}\left(-d\eta^2 +d\rho^2\right)  & {\rm for} & r > r_s, \\
					\frac{16M^2}{r^{}_s} \exp\left\{-\frac{1}{2M} \left[\frac{1}{\sqrt{A_s}}(r-r^{}_s) %1-(2M/r^{}_s)}}
					+r^{}_s\right]\right\}
					\left(-d\eta^2 +d\rho^2\right)   & {\rm for} & r \le r_s,
       	 \end{array}\right. \label{KruShell}
\end{equation}
where $\eta=(v+u)/2$ and $\rho=(v-u)/2$, with $v=e^{V/(2M)} > 0$, $u=-e^{-U/(2M)} < 0$, $V=t+r^*$, $U=t-r^*$, and
$dr^* = dr \sqrt{B(r)/A(r)}$. 
The Kruskal-like coordinates (\ref{KruShell}) are equivalent to radar coordinates (\ref{SchwarzRadar}) up to a conformal transformation (note that $V=(t'+r')/\sqrt{A_o}$ and $U=(t'-r')/\sqrt{A_o}~$ ). 
On the $\eta\rho$-plane, the constant-$t$ hypersurfaces in (\ref{SchwarzShell}) are straight lines
\begin{equation}
  \eta = \frac{1-e^{t/(2M)}}{1+e^{t/(2M)}} \rho,
\end{equation}
and the constant-$r$ hypersurfaces are hyperbolae
\begin{equation}
  \eta^2-\rho^2 = \left\{\begin{array}{lll}
	        \vspace{1mm}-(r-2M)\exp\left\{\frac{r}{2M}\right\} & {\rm for} & r > r_s, \\
					-(r_s-2M)\exp\left\{\frac{1}{2M}\left[\frac{1}{\sqrt{A_s}}(r-r_s)+r_s\right]\right\} & {\rm for} & r \le r_s.
       	 \end{array}\right.
\end{equation}
One can see that the region covered by radar coordinates (\ref{KruShell}) is contained in wedge R in the diagram of the maximal analytic extension of Kruskal coordinates for a Schwarzschild solution of mass $M$ (Figure \ref{SchwS} (left)) and is geodesically complete. Each event in the covered region has a two-way causal connection with the localized observer. In the case of the spherical shell here, there is no room for any counterpart of wedge P in the maximal analytic extension of Rindler coordinates, not to mention any past horizon for the localized observer at $r=r^{}_o$. 

In a static spacetime (\ref{SchwarzShell}), while in (1+1)D, the astronomical distances for a localized observer cannot be determined, one can still formally define the affine distance of an event as the difference of the normalized affine parameter along a null geodesic connecting the event at $r=r^{}_e$ and some point of the worldline of the localized observer at $r^{}_o$ in the future of the event. 
From the geodesic equations, the affine distance reads
$\tilde{r} = \alpha \big|\int^{r^{}_e}_{r^{}_o} \sqrt{AB} dr \big|$ 
up to a constant factor $\alpha$ \cite{HP10, Lin19b}. %160210-1 and the backpage of 190213-1 
We choose $\alpha = 1/\sqrt{A_o}$ %181014-1,3 
to match the radar distance in the neighborhood of the localized observer. Rewriting $r^{}_e$ as $r$, we find the affine distance 
\begin{equation}
	\tilde{r} = \frac{|r-r^{}_o|}{\sqrt{A_o}} \label{daOut}
\end{equation}
for the events outside the spherical shell ($r > r^{}_s$), and
\begin{equation}
	\tilde{r} = \frac{1}{\sqrt{A_o}}
	\left[ r^{}_o-r^{}_s +\sqrt{A_s} (r^{}_s-r)\right]	\label{daIn}
\end{equation}
for the events inside ($r < r^{}_s)$. One can see that $\tilde{r}$ is finite for $r=0$ and $r = 2M$.
The observational coordinates for the observer localized at $r=r^{}_o$ then read 
\begin{equation}
  ds^2 = \left\{\begin{array}{lll}
	        \vspace{1mm} \frac{1}{A_o} \left[-\left(1-\frac{2M}{r^{}_o\pm \tilde{r} \sqrt{A_o}}\right) 
					    d\tilde{\tau}^2 +2d\tilde{\tau}d\tilde{r}\right]  & {\rm for} & r > r_s, \\
					-\frac{A_s}{A_o} d\tilde{\tau}^2 +2 d\tilde{\tau}d\tilde{r}  & {\rm for} & r\le r_s,
       	 \end{array}\right. \label{SchwarzObs}
\end{equation}
where $d\tilde{\tau}=dt'+(d\tilde{r}/A(r))$ for $r>r^{}_s$, $d\tilde{\tau}=dt'+(A_o/A_s)d\tilde{r}$ 
for $r\le r^{}_s$, and $+$ and $-$ correspond to the cases of $r > r^{}_o$ and $r^{}_s < r < r^{}_o$, respectively.
Observational coordinates (\ref{SchwarzObs}) cover almost the same region that radar coordinates (\ref{SchwarzRadar}) do at late times in the Penrose diagram except the neighborhood of past null infinity.
Since the affine distance $\tilde{r}$ in (\ref{daOut}) and (\ref{daIn}) is proportional to $r$, the contours of $\tilde{r}$ in the $\eta\rho$-plane and the Penrose diagram for (\ref{SchwarzObs}) have the same pattern as those of $r$ for (\ref{SchwarzShell}).

If the worldline of the observer is started at some point in past null infinity rather than past timelike infinity, then the situation will be similar to the one with the uniformly accelerated observer in Minkowski space: the spacetime region $R_o$ covered by observational coordinates will be significantly larger than the region  $R_r$ covered by radar coordinates in the Penrose diagram. The border of $R_r$ and $R_o-R_r$ is the past horizon where the radar distance is infinity but the affine distance is finite for the localized observer. 

In Figure \ref{SchwS} (right), we sketch a scenario of a collapsing thin shell similar to the collapsing star in Ref. \cite{DF77} (%rigorous 
calculations can be found in the literature, e.g., Ref. \cite{Po04}). The union of wedges R and F is maximal, and there is no need of attaching wedges L and P or a white hole. 
%Suppose the thin shell contains a few point-like light emitters inside and the shell 
%is semi-transparent. 
%, and an observer is  witnessing the gravitational collapse.
As the shell radius is approaching the Schwarzschild radius ($r_s\to 2M$), an observer localized at $r^{}_o$ outside the shell would perceive that 
%all the images of interior emitters are gradually converging to the images of the shell surfaces 
the thickness of the star in terms of the affine distance is decreasing, since the depth information of different interior points of the star would be suppressed as $\sqrt{A_s}\to 0$ in this limit (note that the $(r_s-r)$ term in (\ref{daIn}) is proportional to $\sqrt{A_s}$). Although that depth information could be resolved in terms of the radar distance whenever $r_s\not= 2M$ (the $(r_s-r)$ term in (\ref{Schwaradar}) is proportional to $1/\sqrt{A_s}$), measuring the radar distances of the shell interior could be much harder than measuring the affine distances in the limit $r_s\to 2M$ because the former needs more historical knowledge about the received signals (echoes) for the observer, e.g., the departure time $\tau_i$ of a radar signal from the observer to the star, which may be lost as $\tau_f-\tau_i$ becomes extremely large in the limit $r_s\to 2M$. Also, the ingoing radar signal may add energy to a nearly black star then turn the star to a black hole. In this case, the radar echo will never come back to the outside world \footnote{Similar observation in (3+1)D in terms of more measures of distance may be relevant to the area law of black hole entropy \cite{Lin19b}.}. Finally, at late times of gravitational collapse, the outside observer will never see the event horizon or the past horizon in the observer's radar coordinates, since no past light cone from the observer will intersect them. The observed horizon at the affine distance $\tilde{r}|_{r^{}_s\to 2M}=\sqrt{r^{}_o(r^{}_o-2M)}$ from the observer is the illusory horizon \cite{HP10}. 
%rather than the past horizon in the observer's radar coordinates or the event horizon.

%%%%%%%%%%%%%%%%%%%%%%%%%%%%%%%%%
\section{Summary and discussion} 
\label{sumdisc}

We have considered the observational coordinates and radar coordinates for the localized observers in (3+1)D Minkowski space in inertial motion (Mi, Section \ref{ALOMink}), in uniform acceleration (Ma, Section \ref{UAO}), and spinning without center-of-mass motion (Ms, Section \ref{spinO}), and for those observers comoving in de Sitter space (dS, Section \ref{deSOO}), fixed at constant radius in (1+1)D Schwarzschild geometry outside of static (Ss) and collapsing (Sc, Section \ref{Schwarz2D}) spherical shells, as well as the cases of non-uniform acceleration in Minkowski space (Mnu, Section \ref{NULA}) and non-eternal inflation (Nei, Section \ref{NEInflat}) where the universe was similar to Minkowski space before the onset of inflation.

\subsection{Regions covered by radar and observational coordinates}
Observational coordinates are determined by a localized observer according to the light or radar signals received. Radar coordinates are determined  with a stronger condition that those received signals are echoes of the radar signals emitted earlier by the same observer. Thus, the spacetime region $R_r$ covered by radar coordinates must be contained by the region $R_o$ covered by the observational coordinates for the same observer. For an observer localized at the origin in Minkowski space, either non-spinning (in Mi) or spinning (in Ms), and for a localized observer fixed in (1+1)D Schwarzschild geometry at a constant radius from the center of a spherical shell (in Ss), 
radar coordinates and observational coordinates for the localized observers at late times appear to cover the same spacetime regions in the Penrose diagrams, where $R_o$ is the closure of $R_r$. The situations are similar in the cases Nei and Sc, 
%of the non-eternal inflation (Nei) and the collapsing star (Sc), 
where neither observational nor radar coordinates can cover the whole universe due to the presence of the event horizon. The region $R_o-R_r$ becomes significant in the Penrose diagrams in Ma and dS.
%for the uniformly accelerated observer in Minkowski space (Ma) and the comoving observer situated at the origin in de Sitter space (dS). 
In these cases, the observers can see through the past horizons of radar coordinates and coordinatize the events 
beyond the reach of radar coordinates with some physical assumptions. 

\subsection{Static limit surface and past horizon}
Coordinate singularities at finite perceived distances in observational coordinates arise in the cases Ma and Ms for non-inertial observers in Minkowski space and dS for a comoving observer in de Sitter space. These coordinate singularities are associated with signature change of the metric component $g^{}_{\tilde{\tau}\tilde{\tau}}$ or $g^{}_{\tilde{t}\tilde{t}}$ in each case and correspond to the static limit surfaces, beyond which no point-like physical object is possible to be seen at rest in the viewpoint of the localized observer. In Ma and dS, the static limit surfaces of observational coordinates coincide with the past horizons of radar coordinates for the same observers. However, in Ms, observational and radar coordinates for the spinning observer share the same static limit surface which is not the past horizon for the observer.

\subsection{Coordinate singularity and acceleration}
While the observers in the cases Ma and Ms are accelerated, and the comoving objects in the case dS look accelerated in the viewpoint of the observer, the accelerations of the observer and/or the comoving objects are not always associated with the coordinate singularities at finite perceived distance in observational coordinates. Indeed, there is no static limit surface at finite perceived distance in the cases Ss and Nei, although in Ss, the localized observer fixed at a constant radius from the center of a static massive spherical shell is at a constant acceleration, 
%or past horizon in radar coordinates  there is no static limit surface at finite perceived distance for the observer, either, though 
and in Nei, the accelerations of the comoving objects become nonzero and never vanish after the onset of inflation. 

\subsection{Event horizon and illusory horizon}
There exist event horizons in the cases Ma, dS, Sc, and Nei. The localized observer at late times in Ma, dS, or Sc would see an illusory horizon \cite{HP10} at some finite distance in her observational coordinates. All the visible physical objects with timelike worldlines passing through the event horizon would appear to approach the illusory horizon and eventually freeze there. In the Penrose diagrams of these cases, an illusory horizon does not have to coincide with the past horizon (in Sc there is even no past horizon). In Nei, only the timelike worldlines passing through the event horizon after the onset of inflation would be observed like that: they would appear to go away from the observer and approach the illusory horizon at late times. Other physical objects would be seen at late times as frozen behind the illusory horizon.  

The non-uniformly accelerated observer in the case Mnu with no event horizon could still see a surface similar to the illusory horizon. 
%in observational coordinates. 
During the period of constant acceleration the visible physical objects also tend to approach that surface and freeze there. However, the spinning observer in Ms with no event horizon could not see any illusory horizon,
%that the physical objects appear to converge to at late times, 
even if a coordinate singularity occurs at finite distance in this case.

%\subsection{Cauchy problems}
%To form a Cauchy problem for a field in the regions covered by observational or radar coordinates, one defines a time coordinate which is timelike in the whole region, so that the dynamical variables on a time slice can represent all those degrees of freedom on past null and timelike infinities. In (Ma) or (dS), while the static limit surface is located at a finite distance in terms of observational coordinates, the surface is still part of the past infinity with respect to the time coordinate $\tilde{t}$ 
%rather than the null coordinate $\tilde{\tau}$. %Note that in (Ms) the static limit surface is not the past infinity.
%Thus a Cauchy problem could be defined within the static limit surface, %For the radar coordinates in (Ma) or (dS), 
%with the information behind the past horizon (in wedge P) which would affect the field in wedge R %can be 
%encoded as the initial data on the past infinity with respect to $\tilde{t}$ or radar time for the localized observer. 
%Using the initial data on the whole past infinity the theory has a full predictability in the whole region covered by radar coordinates for the localized observer at late times. Nothing behind the event horizon can further affect the physics in the visible universe of the observer. Nevertheless, when considering an interacting field theory, past infinity of radar coordinates may not be an asymptotic region where the interactions can be turned off, and so the initial state of the system could be complicated and the concepts from free field theory could not apply. 

\begin{acknowledgments}
I would like to thank Bei-Lok Hu, Kin-Wang Ng, Ue-Li Pen, and Bill Unruh for illuminating discussions.
%on the Cauchy problems and gravitational lensing, and non-eternal inflation. 
I also thank Ya-Zi Wang and Chen-Hau Lee for bringing Refs. \cite{Ki69} and \cite{Bo09} to my attention. This work is supported by the Ministry of Science and Technology of Taiwan under Grant No. MOST 106-2112-M-018-002-MY3 and in part by the National Center for Theoretical Sciences, Taiwan.
\end{acknowledgments}

\end{document}